\documentclass[aps,prb,showpacs,preprintnumbers,twocolumn]{revtex4-2}
\usepackage{amsmath,amssymb}
\usepackage{bm}
\usepackage{tipa}
\usepackage{upgreek}
\usepackage{comment}
\usepackage{mathrsfs}
\usepackage{graphicx}
\usepackage{braket}
\usepackage{enumitem}
\usepackage{mathbbol}
\usepackage{gensymb}
\usepackage[normalem]{ulem}
\usepackage{color}
\usepackage[colorlinks,bookmarks=true,citecolor=blue,linkcolor=red,urlcolor=blue]{hyperref}
\usepackage{hyperref}
\renewcommand{\vec}[1]{\mathbf{#1}}
\usepackage{pifont}

\begin{document}

	\title{Domain wall competition in the Chern insulating regime of twisted bilayer graphene} 
	
	\author{Yves H. Kwan}
	\author{Glenn Wagner}
	\author{Nilotpal Chakraborty}
	\author{Steven H. Simon}
	\author{S.A. Parameswaran}
	\affiliation{Rudolf Peierls Centre for Theoretical Physics, Parks Road, Oxford, OX1 3PU, UK}

	\begin{abstract}
    We consider magic-angle twisted bilayer graphene (TBG) at filling $\nu=+3$, where experiments have observed a robust quantized anomalous Hall effect. This has been attributed to the formation of a valley- and spin-polarized Chern insulating ground state that spontaneously breaks time-reversal symmetry, and is stabilized by a  hexagonal boron nitride (hBN) substrate. We identify three different types of domain wall, and study their properties and energetic selection mechanisms via theoretical arguments and Hartree-Fock calculations adapted to deal with inhomogeneous moir\'e systems. We comment on the implications of these results for transport and scanning probe experiments.
	\end{abstract}
	
 	\maketitle
	
\section{Introduction}
	
Twisted bilayer graphene (TBG) and other moir\'e heterostructures have rapidly emerged as new testbeds for exploring the interplay of strong correlations, superconductivity, and band topology. Early work focused on superconductivity and correlated insulating behaviour~\cite{Cao2018,Cao2018b,Yankowitz1059} near `magic' twist angles at which moir\'e-reconstructed bands near the Fermi energy are nearly dispersionless. More recent transport measurements on near-magic-angle TBG on hexagonal boron nitride (hBN) subtrates have uncovered a zero-field Hall response~\cite{Sharpe605} at filling $\nu=+3$ relative to charge neutrality.  Subsequent experiments have demonstrated the robust quantization of the Hall resistance in units of the von Klitzing constant $h/e^2$~\cite{Serlin} and linked it to the orbital polarization~\cite{tschirhart2020imaging} of the electronic state. 
Such an intrinsic quantized anomalous Hall effect (QAHE) requires time reversal symmetry (TRS) breaking. Theoretical explanations have focused on the hBN substrate, which breaks $C_{2z}$ symmetry via sublattice modulation. This in turn stabilizes an interaction-driven spin and valley-polarized  Chern number $C = \pm 1$ state that spontaneously breaks TRS~\cite{bultinck2020mechanism,Xi2019,zhang2019twisted}.

The emergence of such `orbital Chern insulators' (OCIs) near the magic angle can be roughly understood as follows. 	Absent interactions, there are two nearly-flat bands (each fourfold degenerate due to spin and valley flavors). At charge neutrality, these touch at a pair of Dirac points (DPs), near which the spectrum is similar to monolayer graphene but with a renormalized Fermi velocity \cite{Bistritzer}. However, in hBN-aligned TBG, these DPs are gapped \cite{jung2015origin} due to the substrate-induced sublattice splitting, often leading to bands with $|C| = 1$~\cite{bultinck2020mechanism,zhang2019twisted}. Spontaneous  polarization of a subset of these bands due to interactions can thus give rise to a QAHE. The relevant energetics are  reminiscent of that in quantum Hall ferromagnets (QHFM), but are modified by the presence of a lattice~\cite{QHFMChern}. Detailed Hartree-Fock (HF) studies~\cite{bultinck2020mechanism,LiuHF2019,bultinck2_2019,Xi2019,Yi2020,zhang2019twisted,lin2019,Xie,Cea,Lin} indicate that the valley-polarized state is favoured (though  alternatives have been proposed~\cite{kwan2020excitonic,stefanidis2020excitonic}).
	 
The actual situation encountered experimentally is likely more involved than the intuitive picture painted above. For example, the precise form of the substrate coupling is complicated by the fact that graphene and hBN have a $\sim$2\% lattice mismatch and generically form their own alignment-dependent moir\'e pattern~\cite{Lin,cea2020hBN,macdonald2021}. For small deviations away from a low-order commensurate superstructure, large regions (greater than the TBG moir\'e scale $a_\textrm{M}\sim 14~\text{nm}$) emerge which are distinguished by the local properties of the single-particle bands, in particular the assignment of Chern numbers~\cite{macdonald2021}. Other factors such as lattice imperfections and relaxation can also contribute to the spatial variations. Therefore in the presence of hBN-alignment, we generally expect electron interactions to locally polarize the TBG bands into large domains differing by quantum numbers such a Chern number and valley polarization.

Boundaries between such domains host topologically-protected gapless modes. These domain walls (DWs) are our focus below---note that these are distinct from the helical states that have been observed between  AB/BA stacking regions in an electric field~\cite{huang2018}.
\begin{figure}[!t]
	\includegraphics[width=1\linewidth,clip=true]{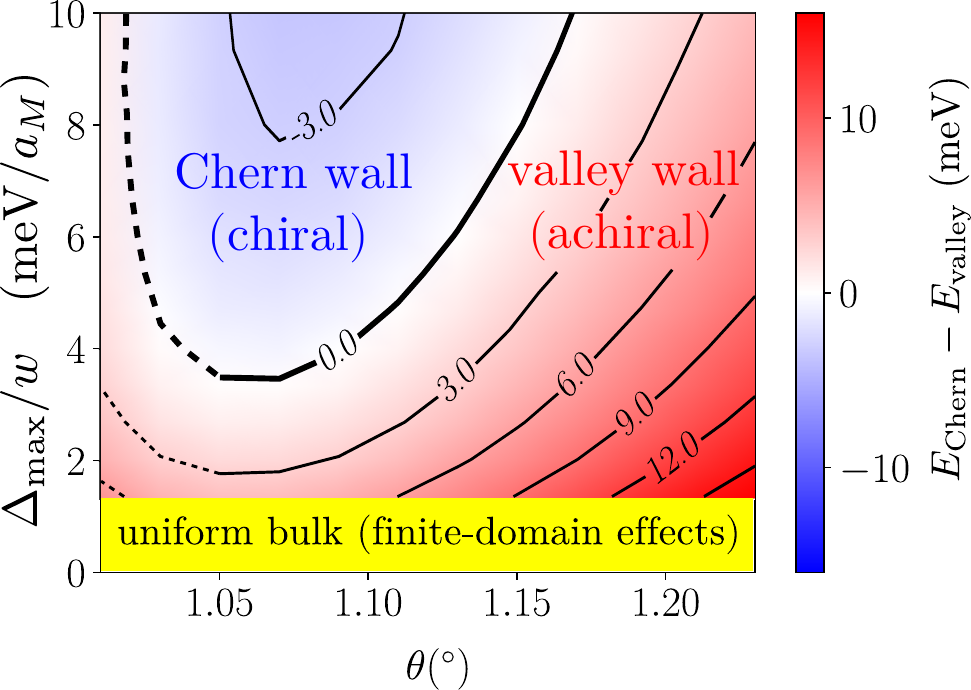}
		\caption{Domain wall phase diagram in the twist angle ($\theta$)-substrate gradient ($\Delta_\textrm{max}/w$) plane, as determined by the energy difference of the two HF solutions $E_\textrm{Chern}-E_\textrm{valley}$ {plotted as contours}. Chern DWs have chiral edge modes whereas valley DWs have achiral modes.
		Valley DWs are favoured at small 
		$\Delta_\textrm{max}/w$, which allows more texturing and hence reduces the exchange penalty. Contours for $\theta<1.05^\circ$ are dashed since the gap to the remote bands becomes small enough that treating the remote bands as inert is a poorer approximation. (We take $N_1=N_2=10,2w=3a_\textrm{M}$.)}

	\label{Phase_diagram}
\end{figure}
Similar DWs in QHFMs have been studied
previously
~\cite{ Falko:1999p1,Mitra:2003p1,NematicValleyPRB2010, Kumar, NematicDWFullTheory, DanonIsingDW} 
and have been  visualized on the surface of bismuth via scanning tunneling microscopy (STM)~\cite{NematicDWExpt}.
As in  QHFMs, DWs in magic-angle TBG generically host a pair of gapless dispersing one-dimensional modes. However, as we show here, in certain parameter regimes 
the edge modes of energetically stable DWs in TBG are  chiral (co-propagating) rather than counter-propagating. Hence universal aspects of chiral Luttinger liquid physics can emerge in this setting, that are absent in QHFM-DWs due to  corrections from interactions between counter-propagating modes~\cite{NematicDWFullTheory}.
    
In this work we present an effective field theory description of the OCI in the presence of a spatially dependent substrate coupling and substantiate this with microscopic self-consistent HF numerics of the DWs. To simplify the analysis, we phenomenologically model the effect of the substrate with a smoothly-varying (on the scale of $a_{\mathrm{M}}$) sublattice splitting $\Delta(\bm{r})$ which can change sign. This captures the $C_{2z}$-breaking effect of the hBN and generally allows for regions with locally different Chern number assignments. In this way, we focus on the interaction physics and interplay between DWs, and remain agnostic to the precise single-particle mechanism that necessitates their formation. The following physical picture emerges between domains with $|C|=1$: There are three types of DWs, which we dub Chern walls (where the Chern number flips across the DW), valley walls (where the valley which is occupied flips across the DW) and intertwined walls (where both the Chern number and valley flip). The intertwined walls are metastable solutions pinned by the moir\'e lattice, whereas the Chern and valley walls are pinned by sign-changing variations in $\Delta(\bm{r})$. The Chern walls are sharp since there is a topological obstruction for forming inter-Chern coherent regions, while the lack of intervalley exchange means that the valley wall is textured. Hence the energetic competition between the Chern and valley walls will be tuned by the gradient of the substrate jump, which tunes the sharpness of the DW (Fig.~\ref{Phase_diagram}).

\section{Effective Field Theory} Assuming full spin polarization, the flat bands of hBN-aligned TBG can be labeled by sublattice ($A, B$) and valley indices ($K, \bar{K})$. We introduce   Pauli matrices $\sigma_\mu (\tau_\mu$) that act in the sublattice (valley) space, with $A, B$ ($K$, $\bar{K}$) corresponding to $\sigma_z =\pm1$ ($\tau_z=\pm1$). (Perfect sublattice-polarization  only emerges in the `chiral limit'~\cite{ChiralLimit}, but it is physically reasonable to retain this label more generally~\cite{khalaf2020charged}.) Before including interactions or substrate effects, bands for each $\tau_z$ eigenvalue meet in a pair of DPs. In this notation, a sublattice potential $\propto \sigma_z$ breaks $C_{2z}$ symmetry, gaps the DPs, and assigns equal and opposite Chern numbers $C = \sigma_z \tau_z$ to the two valleys. At odd integer filling factors, interactions lift the residual  $\tau_z = \pm 1$ degeneracy, leading to an OCI state that spontaneously breaks  valley and  time-reversal symmetries. 
This broken-symmetry state is described by a $CP^3$ nonlinear sigma model (NLSM) that describes fluctuations in the combined sublattice-valley space~\cite{khalaf2020soft}---details are presented in App.~\ref{SecAppNLSM}. However, we can approximate this by a simplified free energy for fluctuations of $Z_2$ and  $O(3)$ order parameters ${m}_z$ {and} $\boldsymbol{n}$:
{We write} 
$F = \int d^2\vec{r} \, f$, where
\begin{equation}\label{eq:FNLSM}
  f\sim \rho_s \left[(\nabla m_z)^2 + \frac{(m_z^2-1)^2} {\beta^2 a_\textrm{M}^2}\right] +{\rho}_v (\nabla \boldsymbol{n})^2 - \frac{\Delta(\vec r)}{a_\textrm{M}^2} m_z n_z.
\end{equation}

{Here $m_z \sim \langle \sigma_z \tau_z \rangle$ represents the local Chern number and $n_z \sim \langle \tau_z \rangle$ is the local valley polarization in} e.g. a 
HF trial state, and we have omitted dynamical terms. However, $n_\mu \sim \langle \sigma_x\tau_\mu\rangle$ for $\mu=x,y$ so that $\boldsymbol{n}$ describes the orientation of the valley order parameter within a single Chern sector, which also entails a rotation between sublattices~\cite{khalaf2020charged,khalaf2020soft}. The second term inside square brackets in~\eqref{eq:FNLSM} suppresses smooth variations of the Chern number.  {The stiffnesses $\rho_s$ and $\rho_v$ should be similar in magnitude, $\beta$ is an anisotropy parameter of order unity, and we have neglected anisotropies for $\boldsymbol{n}$ that are allowed in principle, but expected to be small~\cite{khalaf2020soft}.}
We take the substrate potential $\Delta(\vec{r})$ to be  spatially varying, but sufficiently smooth (on the microscopic lattice scale) to preserve the approximate $U(1)_v$ valley symmetry of TBG~\cite{Bistritzer}.
$F$ is a coarse grained theory valid on scales large compared to $a_M$ and  is not expected to be quantitatively accurate for faster variations. Analysis of the full $CP^3$ theory yields separate $O(3)$ NLSMs for either $\boldsymbol{m} \sim \langle \sigma_x, \sigma_y \tau_z, \sigma_z\tau_x\rangle$ {alone} or $\boldsymbol{n}$ {alone} in low-energy limits with the other held constant (with an easy-axis anisotropy for $\boldsymbol{m}$); however, these cannot be written as combined theory of two coupled $O(3)$ vectors. Instead, we adopt the  description \eqref{eq:FNLSM} in terms of just $m_z$ and $ \boldsymbol{n}$ with the caveat that it is an approximation of a more complete $CP^3$ NLSM  described in App.~\ref{SecAppNLSM}.

We can now identify three distinct types of DW in the OCI. For specificity, we take the DWs to lie parallel to the $y$-axis and separate distinct bulk regions; we fix $m_z =n_z \to +1$ for $x\to -\infty$, e.g. by imposing $\Delta>0$ in this limit. We can then classify DWs based on the limiting behavior of $(m_z, n_z)$ as $x\to \infty$ across the DW:
{\bf(1)} {\it Chern walls}, where the Chern polarization flips at the DW  but the valley polarization stays the same, viz. $(m_z, n_z) \to (-1,1)$ as $x\to\infty$; {\bf(2)} {\it valley walls} where the valley polarization flips while the Chern polarization is unchanged,  $(m_z, n_z) \to (1,-1)$
and {\bf(3)} {\it intertwined walls} where both  Chern and valley indices change so that  $(m_z, n_z) \to (-1,-1)$. Examining $F$, it is evident that sign-changing substrate potentials (corresponding in our example to setting $\Delta<0$ for $x\to\infty$) mandate the existence of intervening  Chern or valley DWs, pinned along lines where $\Delta=0$. 
If $\text{sign}(\Delta(\vec{r}))$ is uniform, then the two order parameters are no longer independent: a DW in ${m}_z$ necessarily induces one in $\boldsymbol{n}$.  One na\"ively expects that in the absence of substrate pinning, such intertwined DWs can  propagate through the sample and annihilate. However, the charge density of the central bands is modulated at the moir\'e scale so DWs experience an interaction-induced periodic potential that may pin their locations even if the substrate is uniform.

\begin{figure}
	\includegraphics[trim={0cm 0cm 0cm 0cm}, width=0.95\linewidth,clip=true]{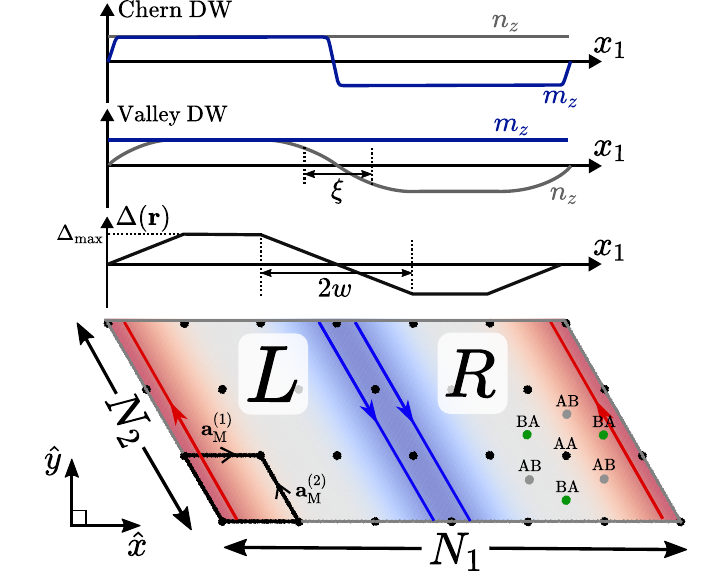}
	\caption{Top-down view of DWs and numerical set-up. We study systems of $N_\text{tot}=N_1N_2$ moir\'e unit cells ($N_1=6,N_2=3$ Chern DW shown for illustration) with periodic boundary conditions. Black dots indicate $AA$ stacking regions, where central bands have largest weight ($AB,BA$ regions are also shown on the right). The substrate potential $\Delta(\bm{r})$ can vary along $\mathbf{a}_\text{M}^{(1)}$ (the sign-changing choice shown can stabilize two Chern or valley walls). Each Chern or intertwined DW hosts two localized co-propagating chiral modes, while a valley wall hosts non-chiral gapless modes. Top plots show the order parameter profiles for the Chern and valley walls.}
	\label{FigSchematic}
\end{figure}

Therefore, for relatively uniform substrates, 
at any finite temperature entropic effects will drive formation of intertwined domains weakly pinned at the moir\'e scale, whose Chern and  valley polarization are locked by the sign of $\Delta$. In contrast, for  non-uniform substrates,
we expect that the sample has Chern or valley domains that track variations in $\text{sign}(\Delta(\vec{r}))$. 
Are Chern or valley DWs  favoured in this  substrate-dominated regime? The answer has  physical implications since Chern (and intertwined) DWs sew together bulk regions which differ in Chern number by $\pm2$. Hence they necessarily host (at least) a pair of chiral co-propagating edge modes---backscattering is forbidden by DW chirality, locking the Luttinger parameter to its non-interacting value. Valley DWs carry counter-propagating modes and their Luttinger parameters can be renormalized by forward scattering, or they can be gapped by backscattering (if microscopic $U_v(1)$-breaking terms which we have neglected here are included). 

Flipping valley polarization $n_z$ across a DW is associated with a loss of exchange energy, penalizing the valley DW relative to the Chern DW. This cost is lowered if valley DWs ``texture'', i.e, rotate smoothly, over some length scale $\xi$ due to the finite valley stiffness. 
We expect that smaller substrate potential gradients will favor valley DWs since they allow more texturing.
Assuming a constant gradient $\Delta_{\text{max}}/w$ that switches sign at $x=0$, variational arguments (App.~\ref{SecAppScaling}) show that 
the width of a valley DW scales as $\xi\sim a_{\textrm{M}}[\rho_v/(a_\textrm{M}\Delta_\textrm{max}/w)]^{1/3}$, (see also~\cite{Mitra_Girvin}). 
In contrast a Chern DW cannot  smoothly interpolate between its two domains (recall $m_z$ in  \eqref{eq:FNLSM} is a $Z_2$ variable). Physically this is because any  `inter-Chern-coherent' region is  topologically forced to admit vortices in its order parameter~\cite{bultinck2020mechanism}, an effect that can counteract valley exchange physics. 
As a result, Chern DWs do not texture, and have a substrate-independent   width $\propto \beta a_M$.

\section{Numerical Methods}
We begin with 
the single-particle band structure of the
continuum model (CM)~\cite{Bistritzer}, with interlayer hopping parameters $w_{AA}=0.08$~eV and $w_{AB}=0.11$~eV~\cite{nam2017,carr2019}.
We innovate a method to treat spatially inhomogeneous substrate configurations: We  project the electronic interactions and the spatially varying substrate potential $\Delta(\bm{r})$ into a fixed `active' Hilbert space spanned by the Bloch functions of the central bands.
This is justified  by the large energy gap between the central  and remote bands. 
We have verified that the central-band Hilbert spaces of the CM solved with $\Delta=0$ have $>99\%$ overlap with those for the values of $\Delta\neq 0$
considered here, justifying this approach. We neglect terms that scatter electrons between valleys, since they are suppressed in twist angle $\theta\ll1$. We use the dual-gate screened Coulomb interaction with Fourier components $V(q)=\frac{e^2}{2\epsilon_0\epsilon_r q}\tanh{qd}$, where  $d=40$~nm is the screening length due to the metallic gates and $\epsilon_r=9.5$ is the relative permittivity~\cite{bultinck2_2019}.

We restrict to substrates that vary only along $\mathbf{a}_\text{M}^{(1)}$ (Fig.~\ref{FigSchematic}), and use the  resulting invariance under $\mathbf{a}_\text{M}^{(2)}$ lattice translations to
focus on HF states that satisfy
\begin{equation}
\langle \hat c^\dagger_{(k_1,k_2)\tau as}\hat c^{\phantom{\dagger}}_{(k_1',k_2')\tau'a's'}\rangle=\delta_{k_2,k_2'}P_{k_1\tau as;k_1'\tau'a's'}(k_2),
\end{equation}
where the HF projector satisfies $\text{Tr}\,P(k_2)=7N_1$ (reflecting the $7$ filled bands) at $\nu=+3$, and $\tau,a,s$ are valley, band, and spin indices. This corresponds to DWs oriented parallel to $\mathbf{a}_\text{M}^{(2)}$.
To avoid double-counting interactions, we subtract a constant projector corresponding to decoupled graphene layers at charge neutrality~\cite{Xie}, which requires us to retain all the remote bands to properly incorporate their screening effects. (We also explored other subtraction schemes.)
We determine the  HF  DW states self-consistently for different substrate configurations. We partition the system along $\mathbf{a}_\text{M}^{(1)}$ into equally sized left ($L$) and right ($R$) regions (Fig.~\ref{FigSchematic}). A convenient reference basis for HF simulations is furnished by the single-particle HF Wannier-Qi (HFWQ) states~\cite{Qi,Scaffidi,Regnault,QiPRX,Liu,bultinck2020mechanism}. These are strip wavefunctions maximally localized perpendicular to the DW, obtained by calculating the bulk HF Bloch orbitals and taking a 1D Fourier transform. We construct the initial HF projector from appropriate combinations of HFWQ states. 
We determine the nature of the converged HF solution by computing  overlaps with the HFWQ orbitals.
We focus on energetically favored fully spin-polarized  states. Further details of the Hamiltonian and numerical modeling can be found in App.~\ref{SecAppContinuum}--\ref{SecAppHFWQ}.

\section{Results}
In order to investigate the competition between Chern and valley DWs, we pick $\Delta(\bm{r})$ so that it takes constant values $\pm\Delta_\text{max}$ deep inside regions $L/R$, and linearly interpolates between them at the boundaries over a width $2w$. Depending on the initial conditions, this leads to a (meta)stable state with two Chern or valley DWs.  Fig.~\ref{Phase_diagram} shows the energy difference between the two configurations as a function of substrate gradient and twist angle. As expected, lower substrate gradients benefit  valley DWs by penalizing valley rotations less. As we show in App.~\ref{SecAppBreakdown}, this manifests in the total exchange energy of valley DWs approaching that of  Chern DWs as the substrate gradient goes to zero. Furthermore, smaller twist angles favour  Chern DWs. 
One or {the} other DW type emerges from this competition as the lowest-energy solution in different parameter regimes.
While beyond-HF corrections (e.g. RPA) and different choices of  double-counting subtraction scheme can modify  precise phase boundaries, our calculations provide strong evidence that Chern-valley DW competition is a \textit{generic} feature of the Chern-insulating regime of TBG. [For small $\Delta_\text{max}$, the  DW cost can exceed the bulk penalty for the `wrong' sign of $m_zn_z$; for finite $N_1$, this  stabilizes a spurious uniform solution  which will become uncompetitive  as $N_{1}\to\infty$.]

\begin{figure}[t!]
	\includegraphics[width=0.9\linewidth,clip=true]{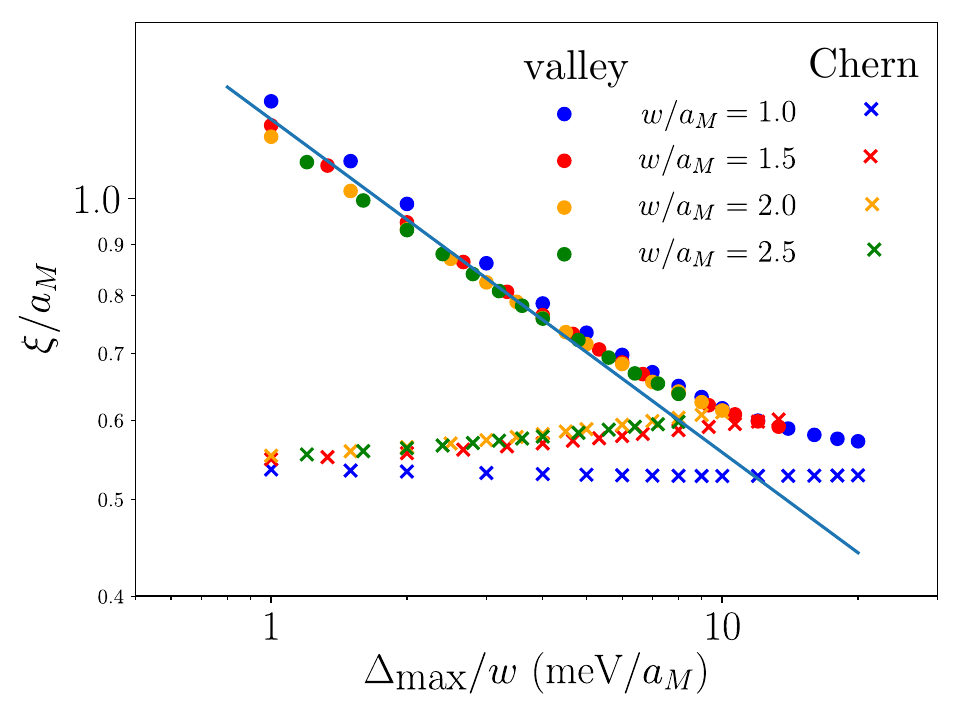}
	\caption{Scaling collapse of width $\xi$ of valley (dots)  and Chern (crosses) DWs against substrate gradient $\Delta_\textrm{max}/w$ at twist angle $\theta=1.2^\circ$, for different widths $w$ of the $\Delta(\vec{r})$ profile. Solid line is the scaling  $\xi\sim a_{\textrm{M}}[\rho_s/(a_\textrm{M}\Delta_\textrm{max}/w)]^{1/3}$ predicted by Eq.~\ref{eq:FNLSM}.  Departures from scaling are seen at large $\Delta_{\text{max}}/w$, where moir\'e-scale physics is relevant.  $N_1=N_2=10$.}
	\label{Scaling}
\end{figure}

Fig.~\ref{Scaling} shows the characteristic width $\xi$ of the Chern and valley DWs for  a range of substrate widths at fixed $\theta=1.2^\circ$. Evidently $\xi$ is controlled by the substrate gradient $\Delta_{\text{max}}/w$. For smaller $\Delta_{\text{max}}/w$, the valley wall is in the scaling regime with an exponent which matches the Ginzburg-Landau prediction of $-1/3$. For large substrates, the scaling regime is cut off by the moir\'e lengthscale. In contrast, the Chern DW is very sharp with $\xi\lesssim a_\text{M}$ independent of $\Delta_{\text{max}}/w$, consistent with a large easy-axis anisotropy $\beta$ in Eq.\eqref{eq:FNLSM} and with the topological arguments~\cite{bultinck2020mechanism} that suppress inter-Chern coherence.

For uniform $\Delta(\bm{r})$ 
intertwined DWs naturally stabilize at half-odd-integer positions between $AA$-stacking regions of maximal central band charge density.
For our largest system sizes and a range of initial conditions, intertwined DWs fails to relax to uniform solutions even after many iterations, instead pinning on the moir\'e-scale charge inhomogeneity. This suggests that such domains  can only grow or shrink in discrete steps of $a_{\mathrm{M}}\sim 14$~nm, i.e. intertwined DWs are long-lived metastable configurations. Despite sharing the same valley exchange physics as valley DWs,  intertwined DWs are far less textured with characteristic length $a_{\mathrm{M}}$, due to Chern number switching (cf. the $m_z n_z$ term in~\eqref{eq:FNLSM}). {With a domain wall tension of $\sim 4\,\text{meV}/a_{\mathrm{M}}$ (see Fig.~\ref{fig:phase_diagram_intertwined} in App.~\ref{SecAppIntertwinedPhase}), this leads to a Curie temperature of order $40\,\textrm{K}$~\cite{Li2014}. Therefore, intertwined wall proliferation is likely not the temperature-limiting factor of the ordered state. In App.~\ref{SecAppChernValley}, we discuss the analogous Ising transition when $\Delta(\bm{r})$ has strong spatial variations and Chern/valley walls are relevant.

\section{Discussion}
We have proposed the existence of three distinct types of DWs in hBN-TBG. Chern and valley DWs  emerge in substrate-fluctuation-dominated samples comprised of domains with opposite local effective sublattice mass.  Their energetic competition depends sensitively on microscopic details such as the twist angle and effective substrate gradient. In contrast, for more uniform substrates the  metastability of intertwined DWs due to moir\'e pinning points to a distinct scenario where valley and orbital physics are locally linked. The substrate modulations that induce Chern/valley DWs also produce additional  bound states at energies far from the Fermi energy (compared to the temperature), absent in the intertwined DW. While non-universal and topologically unprotected, these could provide a way to distinguish the two regimes, e.g. via STM on single-gated samples~(App.~\ref{SecAppSpectral} and \ref{SecAppDWDetails}).

Our results suggest that variations in twist angle and the sign of the substrate potential play a complex role in OCIs. In extreme limits where one type of DW is dominant across the sample, we expect that  they either disrupt global valley order while retaining a robust global magnetization (if valley DWs dominate), or else lead to a vanishing net Chern number but a robust global valley order (if Chern DWs dominate). More generally, we expect a mixed scenario where both situations occur in different regions of a single sample. An added subtlety is that such fluctuations are likely to be quasiperiodic rather than truly random
. In samples where  $\text{sign}(\Delta(\vec{r}))$ is relatively uniform, we expect instead intertwined domains that disrupt both valley order and magnetization, pinned to the moir\'e potential or possibly to  weak variations in $
\Delta(\vec{r})$ or twist angle. Pinning of DWs of various kinds is  likely relevant to understanding current-driven magnetization reversal~\cite{He_MagReversal,huang2020current}. 

Our results have important implications for the observation of the QAHE in TBG and other OCIs. In particular, we provide a candidate explanation for why some TBG samples at filling $\nu=+3$ exhibit the QAHE whereas others do not. If electrical transport is dominated by percolating Chern or intertwined DWs, macroscopic QAH response is always destroyed due to the chiral DW network, which is similar in some respects to a doubled Chalker-Coddington network model~\cite{Chalker_1988}. This would describe a transition between QH plateaus whose Hall conductances differ by $2e^2/h$. Notably, such a transition requires fine-tuning in conventional QH systems but here the relevant network model emerges naturally from disorder. A different situation pertains to  valley DWs which host counter-propagating modes protected against backscattering by $U(1)_v$ conservation. Microscopic $U(1)_v$-breaking terms (neglected above) could open a gap at such achiral DWs and further lower their energy, leading to an orbital ferromagnetic, valley-disordered ground state with a macroscopically robust QAHE -- a `QH random field paramagnet'~\cite{NematicValleyPRB2010}. Intermediate scenarios suggest a new type of network model that has both chiral and  achiral segments. Understanding the properties of these unusually intricate DW networks is an  intriguing open question.

	\textit{Note Added.---}
	During completion of this manuscript, we became aware of a related hybrid Wannier-function scheme for moir\'e systems in Ref.~\cite{hejazi2020hybrid}, which however does not report results for inhomogeneous substrates. Also, recent advances~\cite{kang2020,soejima2020efficient} suggest that some of the results reported here could be within reach of tensor-network techniques, providing a complementary perspective.

	\begin{acknowledgments}  We thank N.~Bultinck, S.L.~Sondhi, C.~Tschirhart,  A.F.~Young, A.~Nahum, and Brian LeRoy for discussions, and  to M.P. Zaletel for a lucid summary of Refs.~\cite{khalaf2020charged} and ~\cite{khalaf2020soft}. We are especially grateful to Allan MacDonald for suggesting we explore different subtraction schemes, which led to quantitatively different results. We acknowledge support from the European Research Council under the European Union Horizon 2020 Research and Innovation Programme via Grant Agreement No. 804213-TMCS (SAP), and from EPSRC Grant EP/S020527/1 (SHS, SAP). Statement  of  compliance  with  EPSRC  policy  framework  on  research  data:  This  publication is theoretical work that does not require supporting research data.
	\end{acknowledgments}
	
{	
\appendix
\onecolumngrid

\section{Continuum model (CM)}\label{SecAppContinuum}
The CM~\cite{Bistritzer} (often referred to as the Bistrizter-MacDonald model) is a widely used approximation to the band structure of TBG. The general premise is that in the low-energy limit, we can focus on momenta near one of the valleys (say valley $K$), and think of four species of fermions (two layers $\times$ two sublattices). We work in the basis of continuum plane waves $\ket{\bm{k},l,\sigma}$ where $\bm{k}$ is measured with respect to global origin of momentum, $l=1,2$ is the layer index, and $\sigma=A,B$ refers to sublattice. The intralayer physics is simple and modelled by (twisted) Dirac cones with Fermi velocity $v_0\simeq9\times10^5$\,ms$^{\text{-1}}$. The interlayer coupling leads to a spatially-modulated hopping amplitude between the layers. In the `dominant harmonic approximation', the CM Hamiltonian is 
	\begin{gather}
		\bra{\bm{k},1}H\ket{\bm{k}',1} = \hbar v_0 \bm{\sigma}^*_{\theta/2}\cdot(\bm{k}-\bm{K}^1)\,\delta_{\bm{k},\bm{k}'}\\
		\bra{\bm{k},2}H\ket{\bm{k}',2} = \hbar v_0 \bm{\sigma}^*_{-\theta/2}\cdot(\bm{k}-\bm{K}^2)\,\delta_{\bm{k},\bm{k}'}\\
		\bra{\bm{k},1}H\ket{\bm{k}',2} = 
		T_1\delta_{\bm{k}-\bm{k}',\bm{0}} + T_2\delta_{\bm{k}-\bm{k}',\mathbf{b}^\text{M}_{1}} + T_3\delta_{\bm{k}-\bm{k}',\mathbf{b}^\text{M}_{2}}\\
		\bm{\sigma}^*_{\theta/2}=e^{-(i\theta/4)\sigma_z}(\sigma_x,\sigma_y^*)e^{(i\theta/4)\sigma_z}\\
		T_1 = \begin{pmatrix}w_{AA}&w_{AB}\\w_{AB}&w_{AA}\end{pmatrix}\\
		T_2 = \begin{pmatrix}w_{AA}&w_{AB}e^{i\phi}\\w_{AB}e^{-i\phi}&w_{AA}\end{pmatrix}\\
		T_3 = \begin{pmatrix}w_{AA}&w_{AB}e^{-i\phi}\\w_{AB}e^{i\phi}&w_{AA}\end{pmatrix}\\
		\phi=\frac{2\pi}{3}
	\end{gather}
where the sublattice degree of freedom has been absorbed into the matrix structure, $\sigma_i$ are the Pauli matrices in sublattice-space, and $\bm{K}^{1,2}$ are the $K$ Dirac point positions rotated by twist angle $\pm\theta/2$. The moir\'e reciprocal lattice vectors (RLVs) for the moir\'e Brillouin zone (mBZ) are $\mathbf{b}_\text{M}^{(1)}=\sqrt{3}k_\theta (\frac{1}{2},\frac{\sqrt{3}}{2})$ and $\mathbf{b}_\text{M}^{(2)}=\sqrt{3}k_\theta (-\frac{1}{2},\frac{\sqrt{3}}{2})$, where $k_\theta=2k_D\sin\theta/2$ is the moir\'e wavevector,  $k_D=4\pi/3\sqrt{3}a$ is the monolayer Dirac momentum, and $a\simeq1.42$\,\r{A} is the C-C bond length. $w_{AA}$ and $w_{AB}$ can be thought of as the interlayer hopping strengths in $AA$ and $AB$ stacked regions. We choose $w_{AA}=0.08$eV and $w_{AB}=0.11$eV to account for corrugation effects~\cite{nam2017,carr2019}. Formally $H$ is an infinite-dimensional matrix in the momentum basis, and a momentum cutoff is required for numerical calculations. For the results in the main text, we keep plane waves within a cutoff that respects TRS and the emergent $D_6$ symmetry of TBG, see Fig.~\ref{SMFigmomentumcutoff}. The equations for valley $K'$ can be deduced by time-reversal.

The CM supplies us with (spin-independent) Bloch functions, which for band $a$ and valley $\tau$ are
\begin{equation}
\psi_{\bm{k}\tau as}(\bm{r})=\frac{e^{i\tau\bm{X}\cdot\bm{r}}}{\sqrt{A}}e^{i\bm{k}\cdot\bm{r}}\sum_{\bm{G}\in\text{RLV},f}e^{i\bm{G}\cdot\bm{r}}c_{\bm{G}\tau a f}(\bm{k})\ket{f,s}\label{EqnCMBloch}
\end{equation}
where $A$ is the system area, $\bm{X}$ is a vector from the absolute zero of momentum to the $\Gamma^\text{M}$-point of the $K$-valley mBZ in the extended zone scheme, $f=(l,\sigma)\in\{1A,1B,2A,2B\}$ is a composite flavor index for layer/sublattice, and $s=\uparrow,\downarrow$ is spin. The corresponding single-particle band energies are $\epsilon^\text{SP}_{\bm{k}\tau a}$. For a finite system with $N_\text{tot}=N_1N_2$ moir\'e unit cells and periodic boundary conditions, the mBZ momenta are discretized as $\bm k=\sum_i \frac{n_i}{N_i} \mathbf{ b}_\text{M}^{(i)}$, where $n_i=0,1,\dots, N_i-1$.

\begin{figure}
	\includegraphics[trim={5cm 3.7cm 5cm 3.7cm}, width=1\linewidth,clip=true]{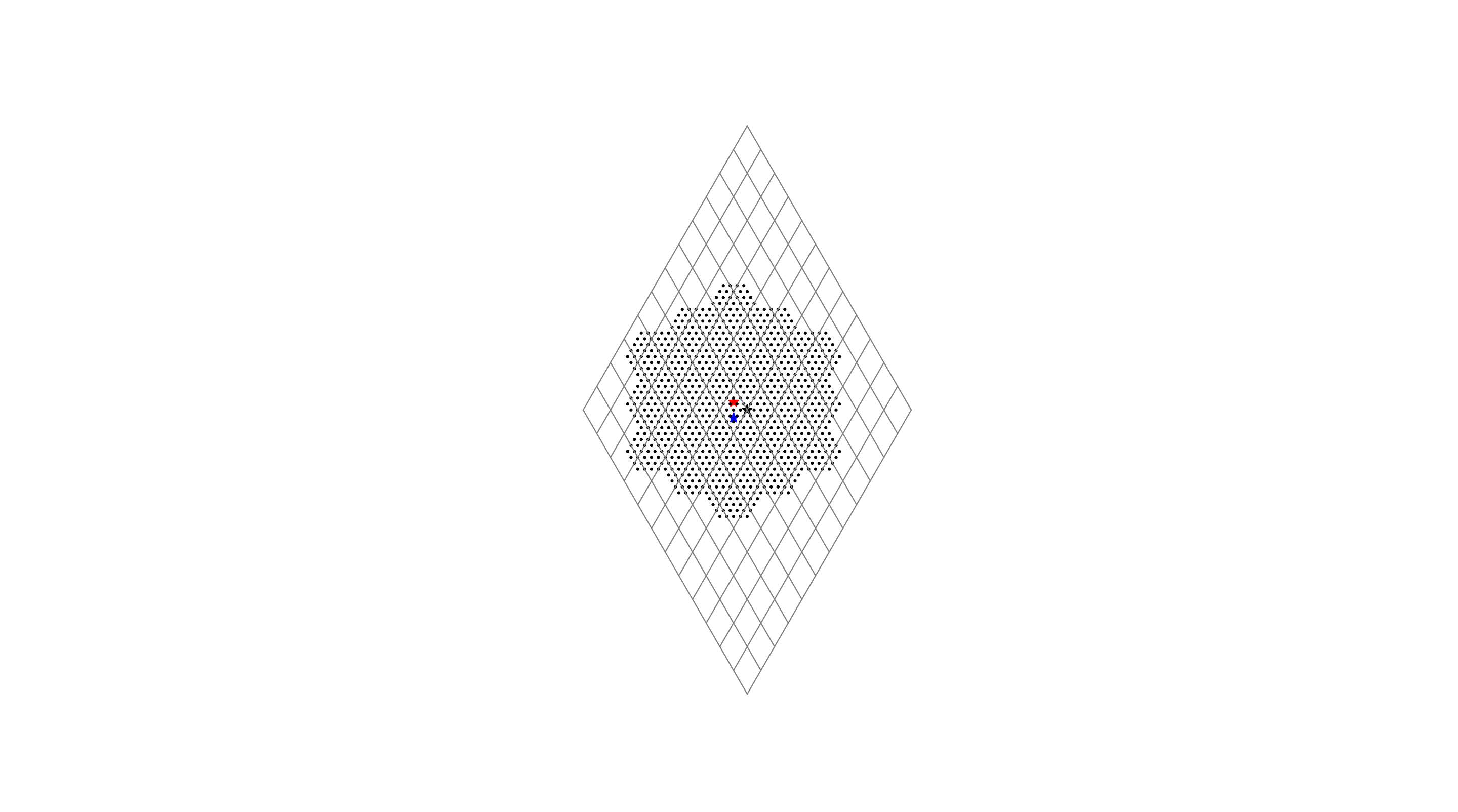}
	\caption{Example symmetric set of plane waves (black dots) that are kept when generating the Bloch functions of the CM. The size of the cutoff shown leads to 228 bands per spin and valley. The red and blue stars are the momenta of the Dirac points of layers 1 and 2. The black star is the $\bm{X}$ vector, which belongs to the $\Gamma^\text{M}$ point of the $K$-valley mBZ. The reference grid indicates plane waves that are equivalent to $\Gamma^\text{M}$. Here we have chosen $N_1=N_2=4$ for illustration, and have only shown the plane waves belonging to layer $1$ and sublattice $A$ in valley $K$. The included plane waves corresponding to the other combinations of layer, sublattice, and valley can be obtained by acting with TRS and the elements of $D_6$.\label{SMFigmomentumcutoff}}
\end{figure}

We define the following form factors that will be useful building blocks for computing various matrix elements
	\begin{equation}
		\lambda_{\bm{G},\tau,a,b}(\bm{k},\bm{k}')
		\equiv\sum_{\bm{G}'\in\text{RLV},f}c^*_{\bm{G}'+\bm{G},\tau af}(\bm{k})c_{\bm{G}'\tau bf}(\bm{k}').
	\end{equation}

\section{Fixed Hilbert space method}\label{SecAppFHS}
The CM can be generalized to include a single-particle sublattice mass (e.g. arising from an aligned hBN substrate) by adding a momentum-diagonal term $\sigma_z$ on one or both of the layers. For instance, adding a constant sublattice mass to one of the layers will gap the Dirac points, leading to Chern bands~\cite{bultinck2020mechanism,zhang2019twisted}. However we are ultimately interested in spatially varying sublattice masses, which is tricky to deal with directly in the CM, especially if projection to the central bands is desired (as is the case here). We address this difficulty by restricting the `active' Hilbert space to the central bands of the substrate-free CM at the outset, and projecting the spatially varying potential into this subspace. Since there is a large energy gap between the central and remote bands at zero substrate, we expect this approximation to the single-particle bandstructure to be justified if the substrate isn't too strong. This can be quantified by comparing the two methods for a uniform substrate---we can either solve the CM model with substrate then project to central bands, or project to central bands first then apply the substrate (fixed Hilbert space). As seen in Fig.~\ref{SMFigFHSbandenergy},\ref{SMFigFHSHilbert}, the two approaches only have small deviations up to the largest experimentally relevant substrate strengths. We note that this is easily generalized to the case where some of the remote bands are additionally included as dynamical.

\begin{figure}
	\includegraphics[trim={0cm 0cm 0cm 0cm}, width=0.6\linewidth,clip=true]{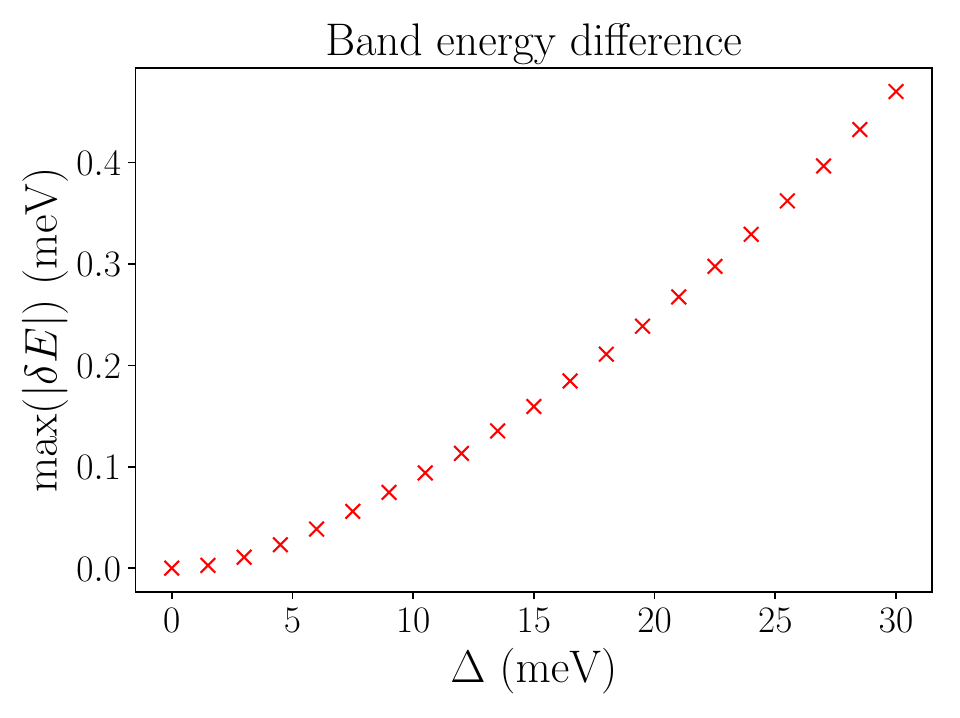}
	\caption{Comparison of the non-interacting band energies between the fixed Hilbert space (FHS) method with external substrate, and the usual continuum model solution with substrate potential. $\text{max}(|\delta E|)$ is defined as the maximum energy difference between the band structures of the two methods over the mBZ.\label{SMFigFHSbandenergy}}
\end{figure}

\begin{figure}
	\includegraphics[trim={0cm 0cm 0cm 0cm}, width=0.6\linewidth,clip=true]{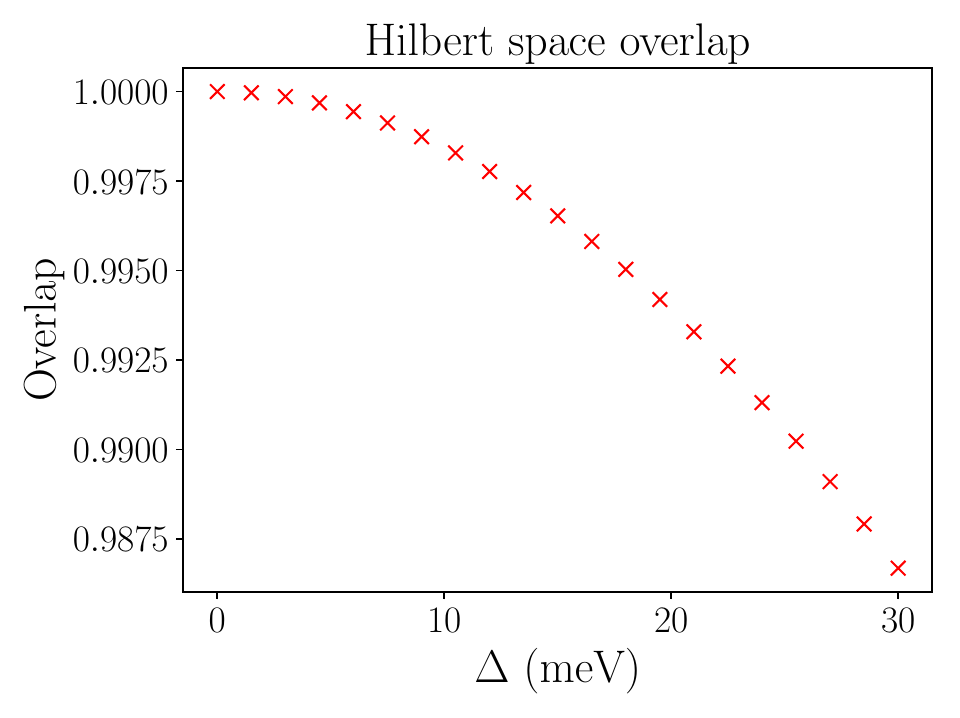}
	\caption{Comparison of the projected Hilbert spaces between the fixed Hilbert space (FHS) method with external substrate, and the usual continuum model solution with substrate potential. Overlap is defined as a normalized $\text{Tr}P_{\text{FHS}}P_{\text{CM}}$, where $P$ is the projector onto the central bands for the respective method. This quantity is 1 if the subspaces are identical.\label{SMFigFHSHilbert}}
\end{figure}

Consider the generic form of the substrate operator (i.e. an effective sublattice splitting) which is local in real-space (which can be expressed as a Fourier series), preserves valley quantum number (valid if the substrate varies smoothly on the microscopic lattice scale), and has the following layer/sublattice structure
\begin{gather}
    \hat{\Delta}(\bm{r})\delta(\bm{r}-\bm{r}')=\bra{\bm{r}}\hat{\Delta}\ket{\bm{r}'},\quad
    \hat{\Delta}(\bm{r})=
    \begin{pmatrix}
    \Delta_1(\bm{r})\sigma_z&0\\
    0&\Delta_2(\bm{r})\sigma_z
    \end{pmatrix},\quad
    \Delta_l(\bm{r})=\sum_{\substack{\bm{q}\in\text{1BZ}\\\bm{Q}\in\text{RLV}}}\Delta_l(\bm{q}+\bm{Q})e^{i(\bm{q}+\bm{Q})\cdot\bm{r}}.
\end{gather}
Matrix elements of the substrate operator between Bloch states can be expressed as
\begin{equation}
    \Delta_{\bm{k}a;\bm{k}'a'}(\tau)=\bra{\psi_{\bm{k}\tau a s}}\hat{\Delta}\ket{\psi_{\bm{k}'\tau'a' s}}=\delta_{\tau\tau'}\sum_{\substack{\bm{G},\bm{G}'\in\text{RLV}\\l,\sigma}}c^*_{\bm{G}\tau a (l\sigma)}(\bm{k})c_{\bm{G}'\tau a' (l\sigma)}(\bm{k}')\Delta_l(\bm{k}-\bm{k}'+\bm{G}-\bm{G}')(\sigma_z)_{\sigma\sigma}.
\end{equation}
where band indices $a,a'$ now only run over the central bands, and the spin index is neglected since the substrate operator is diagonal in spin.

We consider electron interactions that are density-density in valley space, since intervalley exchange is suppressed in $\sim\theta$ (we have checked that inclusion of intervalley exchange does not affect our conclusions). The interaction matrix elements (also independent of spin) can be succinctly expressed in terms of the form factors
\begin{align}
     V^{\bm{k}^\alpha\tau as;\bm{k}^\beta\tau' bs'}_{\bm{k}^\gamma\tau'cs';\bm{k}^\delta\tau ds}&\equiv\bra{\psi_{\bm{k}^\alpha\tau as},\psi_{\bm{k}^\beta\tau' bs'}}\hat{V}\ket{\psi_{\bm{k}^\delta\tau ds},\psi_{\bm{k}^\gamma\tau'cs'}}\\
    &=\frac{1}{A}\sum_{\bm{G}\in\text{RLV}}\tilde{V}(\bm{G}+\lfloor{\bm{k}^\gamma-\bm{k}^\beta}\rfloor)\lambda_{\bm{G}-\lceil{\bm{k}^\alpha+\bm{k}^\beta-\bm{k}^\gamma}\rceil-\lceil{\bm{k}^\gamma-\bm{k}^\beta}\rceil,\tau,a,d}(\bm{k}^\alpha,\lfloor{\bm{k}^\alpha+\bm{k}^\beta-\bm{k}^\gamma}\rfloor)
    \lambda^*_{\bm{G}-\lceil{\bm{k}^\gamma-\bm{k}^\beta}\rceil,\tau',c,b}(\bm{k}^\gamma,\bm{k}^\beta)
\end{align}
where the floor and ceiling notation refer to the mBZ and RLV part of the momentum respectively, and $\tilde{V}$ is the Fourier transform of the interaction potential. We consider both dual-gate screened $\big[\frac{e^2}{2\epsilon_0\epsilon_r q}\tanh{qd}\big]$ and single-gate screened interactions $\big[\frac{e^2}{2\epsilon_0\epsilon_r q}(1-e^{-2qd})\big]$, where  $\epsilon_r=9.5$ and the screening length $d=40$~nm~\cite{bultinck2_2019}. The figures and data in the main text use the dual-gate screened form, but we have verified that using the single-gate screened form does not lead to significant differences.

\section{Hartree-Fock equations}\label{SecAppHF}
If we are considering substrates that conserve $k_2$, then the substrate matrix element can be labelled $\Delta_{k_1a;k_1'a'}(k_2,\tau)$. Since the Hamiltonian conserves $k_2$, we consider the following density matrix/projector
    \begin{equation}
    \langle \hat c^\dagger_{(k_1,k_2)\tau as}\hat c^{\phantom{\dagger}}_{(k_1',k_2')\tau'a's'}\rangle=\delta_{k_2,k_2'}P_{k_1\tau as;k_1'\tau'a's'}(k_2)
    \end{equation}
with $\text{Tr}P(k_2)=7N_1$ for the filling $\nu=+3$. By the standard mean-field decoupling of the interaction term, we obtain the HF effective Hamiltonian
\begin{align}
    \hat{H}^\text{HF}=\sum_{\substack{\tau\tau'ss'\\ab}}\sum_{k_1^\alpha k_1^\beta k_2^\alpha}\mathscr{H}_{k_1^\alpha\tau as;k_1^\beta\tau' bs'}(k_2^\alpha)\hat{c}^\dagger_{(k_1^\alpha,k_2^\alpha)\tau as}\hat{c}_{(k_1^\beta,k_2^\alpha)\tau' bs'}
\end{align}
\begin{align}\label{EqnHFHamiltonianFHS}
    \mathscr{H}_{k_1^\alpha\tau as;k_1^\beta\tau' bs'}(k_2^\alpha)&=\delta_{\tau\tau'}\delta_{ab}\delta_{ss'}\delta_{k_1^\alpha k_1^\beta}\epsilon^\text{SP}_{(k_1^\alpha,k_2^\alpha)\tau a}\\
    &+\delta_{\tau\tau'}\delta_{ss'}\Delta_{k_1^\alpha a;k_1^\beta b}(k_2^\alpha,\tau)\\
    &+\delta_{\tau\tau'}\delta_{ss'}\sum_{\tau''s''}\sum_{\substack{cd\\k_1^\gamma k_1^\delta k_2^\beta}}V^{(k_1^\alpha,k_2^\alpha)\tau as;(k_1^\gamma,k_2^\beta)\tau''cs''}_{(k_1^\delta,k_2^\beta)\tau''ds'';(k_1^\beta,k_2^\alpha)\tau bs}P_{k_1^\gamma \tau''cs'';k_1^\delta \tau''ds''}(k_2^\beta)\\
    &-\sum_{\substack{cd\\k_1^\gamma k_1^\delta k_2^\beta}}V^{(k_1^\alpha,k_2^\alpha)\tau as;(k_1^\gamma,k_2^\beta)\tau'cs'}_{(k_1^\beta,k_2^\alpha)\tau'bs';(k_1^\delta,k_2^\beta)\tau ds}P_{k_1^\gamma \tau'cs';k_1^\delta \tau ds}(k_2^\beta)\\
    &+\delta_{\tau\tau'}\delta_{ss'}\delta_{k_1^\alpha k_1^\beta}H^{\text{scr}}_{(k_1^\alpha,k_2^\alpha)\tau ab}.
\end{align}
The last term above accounts for the double-counting of interactions~\cite{LiuHF2019,Xie} and the screening from remote filled valence bands. It takes the form 
\begin{align}
\hat{H}^\text{scr}=&\sum_{\bm{k}\tau abs}\bigg[\sum_{\tau''s''}\sum_{\substack{m\\\bm{k}'}}V^{\bm{k}\tau a;\bm{k}'\tau''m}_{\bm{k}'\tau''m;\bm{k}\tau b}-\sum_{\substack{m\\\bm{k}'}}V^{\bm{k}\tau a;\bm{k}'\tau m}_{\bm{k}\tau b;\bm{k}'\tau m}
\\
&-\sum_{\tau''s''}\sum_{\substack{xy\\\bm{k}'}}V^{\bm{k}\tau a;\bm{k}'\tau''x}_{\bm{k}'\tau''y;\bm{k}\tau b}P^0_{xy}(\bm{k}',\tau'')+\sum_{\substack{xy\\\bm{k}'}}V^{\bm{k}\tau a;\bm{k}'\tau x}_{\bm{k}\tau b;\bm{k}'\tau y}P^0_{xy}(\bm{k}',\tau)\bigg]\hat{c}^\dagger_{\bm{k}\tau as}\hat{c}^{\phantom{\dagger}}_{\bm{k}\tau bs}
\end{align} 
where $m$ runs over all valence bands below the central bands and are assumed to be fully-filled, and $x,y$ run over all bands. $P^0$ is a reference projector (in the CM basis), which we take in the main text to be the density of decoupled graphene layers at charge neutrality~\cite{Xie}\footnote{We thank Allan MacDonald for suggesting this choice of reference projector.}. This choice is physically motivated since the parameters (such as the Dirac velocity) of the CM model are taken from ab initio calculations of monolayer graphene, and we should therefore only count interactions about this reference point. In the calculation of $H^\text{scr}$, we retain all bands of the CM (up to the plane wave cutoff). Alternative subtraction schemes have been proposed based on choosing a reference filling $\nu_0$ of the CM bands~\cite{LiuHF2019,Yi2020,hejazi2020hybrid}. Note that in this case, the contributions from remote bands in $\hat{H}^\text{scr}$ completely cancel. We show the phase diagram for the energy difference between Chern and valley DWs for the $\nu_0=+4$ subtraction scheme in Fig.~\ref{fig:phase_diagram_full}. The results for the two different subtraction schemes show the same qualitative behaviour in that the valley wall is more energetically favourable at small substrate gradients and at large twist angles. However, quantitative differences mean that the Chern wall is favoured in a larger region of the phase diagram for the $\nu_0=+4$ subtraction scheme compared to the decoupled layers subtraction scheme.

We find that spin is maximally polarized for all of our self-consistent HF solutions. However we cannot neglect spin completely because, depending on the subtraction scheme, filled spin bands can still contribute to the moir\'e scale charge distribution.

\begin{figure}
	\includegraphics[trim={0cm 0cm 0cm 0cm}, width=0.6\linewidth,clip=true]{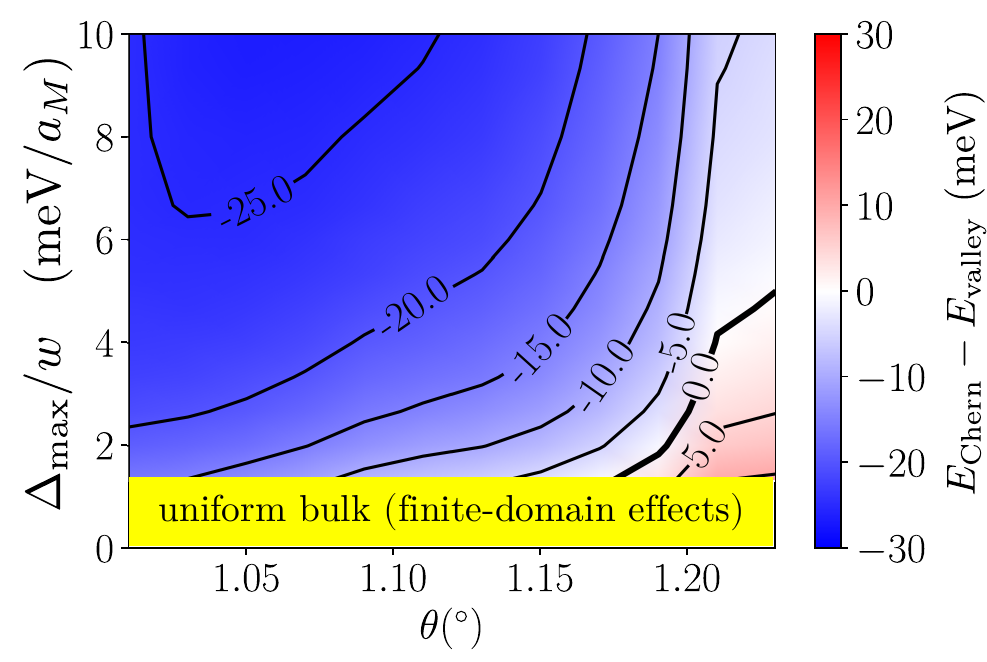}
	\caption{Phase diagram of the energy difference between Chern wall and valley wall, $E_\textrm{Chern}-E_\textrm{valley}$. This is the same as Fig.~1 in the main text, however here we use the subtraction scheme for the interactions with filled central bands, i.e.~the reference projector has $\nu_0=+4.$ $N_1=N_2=10,2w=3a_\textrm{M}$.}
	\label{fig:phase_diagram_full}
\end{figure}

 \section{Hartree-Fock Wannier-Qi orbitals}\label{SecAppHFWQ}

Consider solving the fixed Hilbert space model for a fixed uniform substrate $\Delta$. Since the bulk states are fully valley-polarized, we can assume we are discussing a band with a particular valley and drop the valley index below. Similarly the dependence of various quantities (such as the HF ground state) on $\Delta$ will be kept implicit. For a uniform substrate, the HF solution will conserve crystal momentum in both directions. Therefore, a particular HF band (say the unfilled orbitals at $\nu=+3$) can be described by the coefficients $c^\textrm{HF}_{\bm{G}\tau af}(\bm{k})$ of the plane wave expansion, similarly to how the CM Bloch functions were defined in Eqn.~\eqref{EqnCMBloch}. Note that $a$ now labels a particular HF band. Of course the values of the coefficients will in general differ since the interaction will mix the valence and conduction bands of the non-interacting CM.

We pick the periodic (but not necessarily smooth) gauge, i.e.
\begin{equation}
    c^\textrm{HF}_{\bm{G}\tau af}(\bm{k}+\mathbf{b}_\text{M}^{(i)})=c^\textrm{HF}_{\bm{G}+\mathbf{b}_\text{M}^{(i)}\tau af}(\bm{k}).
\end{equation}
The discretized version of the Berry connection is given by
\begin{equation}
    \mathcal{A}_{\tau a}^{(1)}(n_1,n_2)=\mathfrak{Im} \log \sum_{\bm{G},f} c^\textrm{HF*}_{\bm{G}\tau af}(n_1,n_2)c^\textrm{HF}_{\bm{G}\tau af}(n_1+1,n_2)
\end{equation}
and similarly for $\mathcal{A}_{\tau a}^{(2)}(n_1,n_2)$, where we have parameterized the mBZ momentum in terms of the integers $n_1,n_2$: $\bm k=\sum_i \frac{n_i}{N_i} \mathbf b_\text{M}^{(i)}$. The Hartree-Fock Wannier-Qi (HFWQ) functions are given by \cite{Scaffidi,Qi}
\begin{equation}
\label{eq:Bloch_to_WQ_SM}
        \ket{W_{K\tau as}}=\sum_{n_1=0}^{N_1-1}f_{K\tau a}(n_1)\ket{\psi^\textrm{HF}_{\bm{k}\tau as}},
\end{equation}
where $K=\frac{n_2}{N_2}+\tilde n_1$, and $\tilde n_1=0,1,\dots,N_1-1$. Here $\ket{\psi^\textrm{HF}_{\bm{k}\tau as}}$ refers to the HF orbital of the HF band under consideration. $K$ can take $N_\textrm{tot}$ different values, consistent with the fact that we started off with $N_\textrm{tot}$ states. $0\leq K<N_1$ is also approximately the average position of the HFWQ wavefunction in the $\mathbf a_\text{M}^{(1)}$ direction (it would be exactly the average position if the Berry curvature were uniform). The coefficients in the expansion above are
\begin{equation}\label{EqnWQfcoefficient}
f_{K\tau a}(n_1)=\frac{1}{\sqrt{N_1}} e^{-\mathrm{i}\sum_{n_1'=0}^{n_1} \mathcal{A}^{(1)}_{\tau a}\left(n_1',n_2\right)-\mathrm{i} \frac{2\pi n_1}{N_1}\left(\tilde n_1-\frac{\theta_{\tau a}\left(n_2\right)}{2 \pi}\right)},
\end{equation}
where $\theta_{\tau a}\left(n_2\right)=\bigg(\sum_{n_1=0}^{N_1-1} \mathcal{A}^{(1)}_{\tau a}\left(n_1,n_2\right)\bigg)\mod 2\pi$. The Chern number can be obtained by inspecting the winding of the 1D-polarization $\theta_{\tau a}\left(n_2\right)$.

\section{Non-linear Sigma Model\label{SecAppNLSM} Description of Domain Walls}
The effective model is described in terms of a basis of eight states corresponding to sublattice ($\sigma=A,B$), valley ($\tau=K,\bar{K}$), and spin ($s=\uparrow,\downarrow$) degrees of freedom. We let $\alpha,\beta$ be combined indices. The states are naturally divided into Chern sectors $C=\sigma_z\tau_z$. Consider starting from a uniform insulating/semimetallic Slater determinant state at filling $\nu$ described by projector $P(\bm{k})$. The energy of long-wavelength fluctuations of this state can be described using the non-linear sigma model (NLSM) derived in Refs.~\cite{khalaf2020charged,khalaf2020soft} in terms of the matrix-valued field $Q=2P-1$
\begin{gather}
    \label{SMEqnNLSM}E\left[\tilde{Q}(\bm{r})\right]=\frac{\rho}{8}\text{tr}(\nabla\tilde{Q})^2-\frac{\alpha}{4}\text{tr}(\tilde{Q}\gamma_z)^2+\frac{J}{8}\text{tr}[(\tilde{Q}\gamma_x)^2+(\tilde{Q}\gamma_y)^2]-\frac{\lambda}{8}\text{tr}[(\tilde{Q}\gamma_x\eta_z)^2+(\tilde{Q}\gamma_y\eta_z)^2]-\frac{\Delta(\bm{r})}{2}\text{tr}(\tilde{Q}\gamma_z\eta_z)\\
    Q_{\alpha\beta}(\bm{k})=\langle[c^\dagger_{\bm{k}\beta},c_{\bm{k}\alpha}]\rangle,\quad Q(\bm{k})^2=1,\quad\text{tr}Q(\bm{k})=2\nu\\
    \gamma_{x,y,z}=(\sigma_x,\sigma_y\tau_z,\sigma_z\tau_z),\quad \eta_{x,y,z}=(\sigma_x\tau_x,\sigma_x\tau_y,\tau_z).
\end{gather}
The dynamical term has been omitted above. $\tilde{Q}$ is related to $Q$ by a $\bm{k}$-dependent transformation, which accounts for the momentum-space vortices when performing inter-Chern rotations~\cite{khalaf2020soft}. Ground states therefore involve a $\bm{k}$-independent uniform $\tilde{Q}$. The first term of Eqn~\ref{SMEqnNLSM} is a gradient cost derived in the $U(8)$-symmetric limit (this can be done exactly if the Berry curvature is concentrated at a point). The non-symmetric terms in the Hamiltonian are incorporated primarily as mass terms, but they can also make the stiffness anisotropic in the $\tilde{Q}$-manifold. The $\alpha$-term reflects the energetic penalty of inter-Chern coherence due to the vortex lattice~\cite{bultinck2020mechanism,khalaf2020soft}. The $J$-term arises from the finite dispersion of the central bands, while $\lambda$ parameterizes the part of the interactions that break the $U(4)\times U(4)$ symmetry. To account for the external sublattice potential, we have added a linear term with coefficient $\Delta(\bm{r})$, which acts as an effective sublattice mass.

We consider the spinless problem for simplicity, and work at $\nu=-1$ (this corresponds to one filled central band---the discussion for $\nu=1$ is analogous since it involves specifying one empty band). This allows us to define a $CP^3$ field via $d^\dagger_{\bm{k}}=\sum_\alpha w_\alpha c^\dagger_{\bm{k}\alpha}$ where $d$ is the operator for the filled band. This leads to 
    \begin{equation}
        Q_{\alpha\beta}=2w_\alpha w^*_\beta-\delta_{\alpha\beta}
    \end{equation}

\begin{table}[]
\begin{tabular}{c|c|c|c|c}
Basis State & $\sigma_z=\gamma_z\eta_z$ & $\tau_z=\eta_z$ & $C=\gamma_z$ & $CP^3$ field $w$ \\ \hline\hline
$KA$ & $+1$ & $+1$ & $+1$ & $(1,0,0,0)$ \\ \hline
$\bar{K}A$ & $+1$ & $-1$ & $-1$ & $(0,1,0,0)$ \\ \hline
$KB$ & $-1$ & $+1$ & $-1$ & $(0,0,1,0)$ \\ \hline
$\bar{K}B$ & $-1$ & $-1$ & $+1$ & $(0,0,0,1)$ \\
\end{tabular}\label{SMTabBasis}\caption{}
\end{table}
    
We now show how the NLSM may reduce to a $CP^1$ theory when discussing domain walls in the large $\alpha$ limit. Table~\ref{SMTabBasis} shows the properties of the basis states. We consider each configuration in turn. While $\alpha$ may not actually be much larger than $J,\lambda$~\cite{khalaf2020soft}, considering a dominant $\alpha$-term is a useful organizing principle since it emphasizes the (spinless) $U(2)\times U(2)$ division into Chern sectors.

\begin{enumerate}
\item \textbf{Uniform solution}: Consider first the case of zero sublattice mass. The $\alpha$-term forces us to choose a Chern sector $\gamma_z=\pm1$ to place our filled band in. Within a given Chern sector, we are left with an effective $CP^1$ field. This reflects the fact that $CP^1_{\gamma_z=1}\times CP^1_{\gamma_z=-1}$ can be embedded into $CP^3$. For instance for $\gamma_z=1$, we have $w=(\cos\frac{{\theta}}{2},0,0,e^{i\phi}\sin\frac{\theta}{2})$. The $J,\lambda$ terms vanish for this set of states. Hence at this order in the field theory, the uniform QAH ($\theta=0,\pi$) and intervalley coherent states (in-plane) are degenerate.  A finite sublattice mass (whether external or dynamically generated) will break this degeneracy by selecting the state with the correct sublattice polarization. Hence there are two degenerate ground states labelled by the Chern number $\gamma_z$.
\item \textbf{Valley wall}: A sign-changing substrate $\sim\sigma_z$ means that we are forced to rotate from $\{KA,\bar{K}A\}$ to $\{KB,\bar{K}B\}$ between the bulks. To satisfy the $\alpha$-term, we should only rotate within the same Chern sector $\gamma_z$---this automatically means we switch valley across the wall. Hence we are again left with an effective $CP^1$ field (with the same $CP^1_{\gamma_z=1}\times CP^1_{\gamma_z=-1}$ embedding as in the discussion for the uniform solution). The $J,\lambda$ terms do not contribute. Therefore the $CP^1$ theory (which can be recast into a unit 3-vector $\bm{n}=\langle\bm{\eta}\rangle$) involves just the stiffness term and the out-of-plane sublattice term
\begin{equation}
E[\bm{n}(\bm{r})]=\frac{\rho_v}{4}(\nabla\bm{n})^2-\alpha-\gamma_z \Delta(\bm{r})n_z.    
\end{equation}

\item \textbf{Chern wall}: A sign-changing substrate $\sim\sigma_z$ means that we are forced to rotate from $\{KA,\bar{K}A\}$ to $\{KB,\bar{K}B\}$ between the bulks. This time the $\alpha$-term cannot be fully satisfied because by definition the Chern wall rotates between Chern sectors. Consider for concreteness that the left and right bulks are in $KA$ and $KB$ respectively. The most natural rotation is within the $\{KA,KB\}$ subspace, which would allow for a $CP^1$ description. In principle nothing stops the system from rotating into the other valley as well. However this would be energetically disadvantageous due to the stiffness term---if the $J,\lambda$ are small then the minimal configuration should involve traversing the shortest path between $KA$ and $KB$ to reduce the gradient cost. In this case, $\tilde{Q}$ commutes with $\eta_z$, so that the $J$ and $\lambda$ terms act to largely cancel each other out. In this limit, the $CP^1$ theory (which can be recast into a unit 3-vector $\bm{m}=\langle\bm{\gamma}\rangle$) involves the stiffness term, the easy axis $\alpha$-term, and the substrate term
\begin{equation}
E[\bm{m}(\bm{r})]=\frac{\rho_s}{4}(\nabla\bm{m})^2-\alpha m_z^2+\frac{\lambda-J}{2}(m_z^2-1)-\eta_z \Delta(\bm{r})m_z.    
\end{equation}
Note that this involves a different $CP^1_{\eta_z=1}\times CP^1_{\eta_z=-1}$ embedding than the valley wall. We have used a different stiffness $\rho_s$ than the valley wall since this is generically allowed by the symmetries. 

Comparing the effective $CP^1$ theories for the Chern wall and the valley wall, the only difference (ignoring the $J-\lambda$ contribution) is that the easy-axis anisotropy term of the Chern wall is frustrated with the stiffness term, while the anistropy term of the valley wall is always satisfied. This would seem to suggest that the valley wall should always be energetically better than the Chern wall. However what this simple analysis does not capture is that the Chern wall does not actually rotate from complete $A$ to $B$ sublattice polarization. If it did, then the exchange physics would be the same as the valley wall (since the interaction is density-density in layer, sublattice, spin and valley space). But the actual HF bulks for the Chern wall have finite overlap due to partial sublattice polarization, and hence finite inter-domain exchange. This counteracts the anisotropy cost, and helps facilitate Chern-valley DW competition.

\item \textbf{Intertwined wall}: Since the sublattice mass is of constant sign, the bulks have the same fixed value of $\sigma_z$. However for this class of solutions, we require the system to rotate between valleys. Similar considerations to the Chern wall lead to yet another $CP^1_{\sigma_z=1}\times CP^1_{\sigma_z=-1}$ embedding, where here each $CP^1$ is restricted to a fixed sublattice, so that the 3-vector can be given by $\bm{t}=\langle\bm{\tau}\rangle$
\begin{equation}
E[\bm{t}(\bm{r})]=\frac{\rho_t}{4}(\nabla\bm{t})^2-\alpha t_z^2-\sigma_z \Delta(\bm{r}).    
\end{equation}

\end{enumerate}

\begin{figure}[h!]
	\includegraphics[width=0.8\linewidth,clip=true]{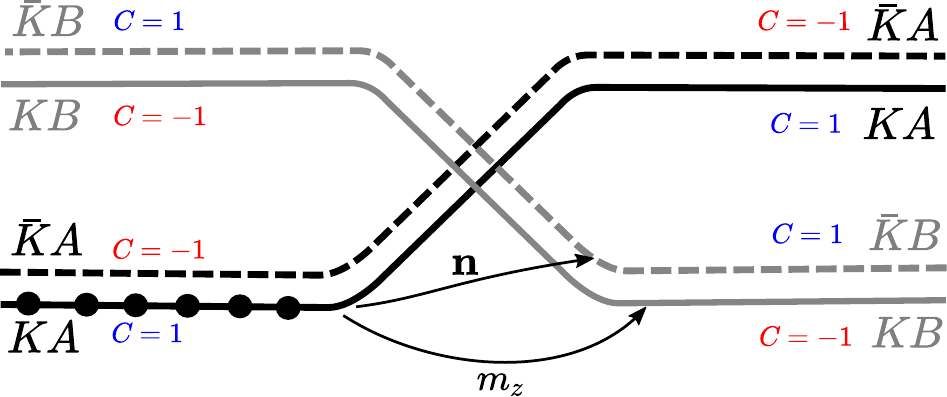}
	\caption{Schematic of energy levels for a sign-changing substrate $\Delta(x)$. Consider the spinless theory at $\nu=-1$. On the left, we assume that we are polarized in $KA$. A valley wall involves a Chern-filtered valley rotation from $KA$ to $\bar{K}B$, which is captured by the Ginzburg-Landau order parameter $\bm{n}$. A Chern wall has global valley polarization, but swaps between sublattices (and hence Chern sectors). This can be captured by an effective Ising degree of freedom $m_z$.} 
	\label{SMFigFourbands}
\end{figure}

\subsection{Ginzburg-Landau Theory}
While the different $CP^1$ theories provide adequate descriptions of each individual DW, it is instructive to construct an approximate Ginzburg-Landau theory that can incorporate multiple DWs into a single formulation. We focus on the valley and Chern walls that were described by $O(3)$ vectors $\langle\bm{\eta}\rangle$ and $\langle\bm{\gamma}\rangle$. Consider a sign-changing substrate that splits the states as shown in~Fig~\ref{SMFigFourbands}. Starting in $KA$ on the left, the valley wall wants to end up in $\bar{K}B$ while the Chern wall wants to end up in $KB$. The intra-Chern rotation of the valley wall can again be described by the vector $\bm{n}\sim\langle\bm{\eta}\rangle$, which has the interpretation of a Chern-filtered valley polarization. However due to the strong $\alpha$-term, the field $\langle\bm{\gamma}\rangle$, which involves inter-Chern rotations, can be approximated as an Ising field $m_z$. This can be roughly thought of as the order parameter corresponding to the Chern number. The substrate term can then be summarized as the coupling $\Delta(\bm{r})n_zm_z$ between the order parameters. A phenomenological Ginzburg-Landau free energy density that accounts for the stiffnesses, easy-axis anisotropy, and substrate coupling is
\begin{equation}\label{eq:AppFNLSM}
  f\sim \rho_s \left[(\nabla m_z)^2 + \frac{(m_z^2-1)^2} {\beta^2 a_\textrm{M}^2}\right] +{\rho}_v (\nabla \boldsymbol{n})^2 - \frac{\Delta(\vec r)}{a_\textrm{M}^2} m_z n_z
\end{equation}
as shown in the main text. Various parameters have been rescaled by the moir\'e length.

\section{Domain Wall Scaling}\label{SecAppScaling}
To address the valley DW texturing described in the main text, we can analyze the Ginzburg-Landau theory in more detail. Taking a constant $m_z=1$,
\begin{equation}
    F\sim \int_{x,y}\rho_v (\nabla \boldsymbol{n})^2 + \frac{\Delta(\vec r)}{a_\textrm{M}^2}n_z-\beta_v n_z^2,
\end{equation}
where we have added an anisotropy term $\beta_v$ for generality. In the limit where the anisotropy is negligible, the width of the DW will be $\xi\sim a_{\textrm{M}}(\rho_v/\Delta(x=a_{\textrm{M}}))^{1/3}$. To show this, let us focus on the case where we have translational invariance in the $y$-direction and let us consider an ansatz for a DW which has a length scale $\xi$ such as $n_z(x)\sim \tanh(x/\xi)$. Assume that $\Delta(x)=\frac{\Delta_\textrm{max}}{w}x$ in the entire range where the DW is textured, i.~e.~ $w\gg\xi$. The energy difference of the valley wall solution compared to the uniform solution $n_z(x)=1$ with substrate $\Delta(x)=\frac{\Delta_\textrm{max}}{w}|x|$ evaluates to 
\begin{equation}
    \Delta F(\xi)/L_2
    \sim \frac{\rho_v}{\xi} + \frac{\Delta_\textrm{max}}{w}\frac{\xi^2}{a_\textrm{M}^2}+\beta_v\xi,
\end{equation}
where we are neglecting dimensionless quantities of order unity. Setting $\partial_\xi\Delta F(\xi)=0$, we find
\begin{equation}
    0\sim-\frac{\rho_v}{\xi^2}+\frac{\Delta_\textrm{max}}{w}\frac{\xi}{a_\textrm{M}^2}+\beta_v
\end{equation}
Neglecting the anisotropy term, we find
\begin{equation}
    \frac{\xi}{a_M}\sim\bigg(\frac{\rho_v}{\frac{\Delta_\textrm{max}}{w/a_M}}\bigg)^{1/3}
\end{equation}
If we neglect the substrate gradient term, we find
\begin{equation}
    \xi\sim\sqrt{\frac{\rho_v}{\beta_v}}
\end{equation}
The crossover between the substrate dominated and the anisotropy dominated regimes occurs when 
\begin{equation}
    \frac{\Delta_\textrm{max}}{w/a_M}\sim \frac{(\beta_v a_M^2)^{3/2}}{\sqrt{\rho_v}}
\end{equation}
The scaling behaviour is only strictly valid in the limit $\xi\lesssim w$ (so that we can assume the linear form of the substrate throughout the region where the DW is textured) and $\xi\gtrsim a_M$ (so that the moir\'e scale physics doesn't determine the size of our domain wall). 
\section{Breakdown of domain wall energies as a function of substrate}\label{SecAppBreakdown}
In Fig.~\ref{FigEDW} we show the energy difference between the Chern and valley DWs as a function of substrate gradient and we also show the difference in exchange and direct energy. The Chern wall incurs a large Hartree penalty, whereas the valley wall incurs a large exchange penalty due to the lack of intervalley exchange. As the substrate gradient is lowered, the valley wall textures more, thus reducing the exchange penalty. 

	\begin{figure}
	\includegraphics[trim={0cm 0cm 0cm 0cm}, width=0.5\linewidth,clip=true]{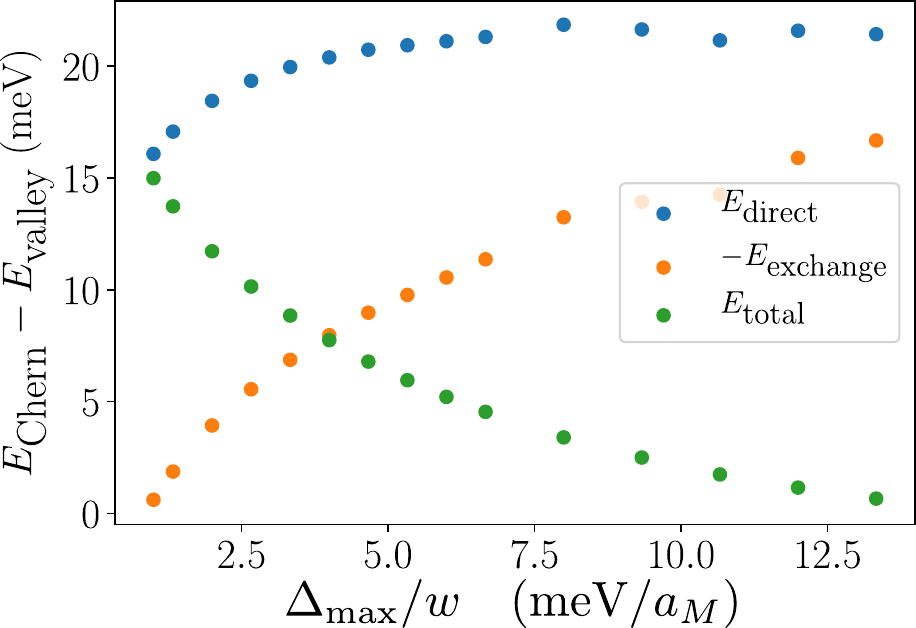}
	\caption{Breakdown of the energy difference between the Chern wall and valley wall into a direct and exchange part. We do not show the single-particle contributions to the energy since these are comparable for the two DW types. $\theta=1.2^\circ$ and $N_1=N_2=10$.}
	\label{FigEDW}
\end{figure}

\section{Intertwined wall phase diagram}\label{SecAppIntertwinedPhase}
Fig.~\ref{fig:phase_diagram_intertwined} shows the phase diagram of the energy difference between intertwined wall and uniform solution. The substrate potential $\Delta(\bm{r})$ is the same as that for the Chern and valley walls, except the modulus is taken. There is always an energy penalty for having an intertwined DW due to the loss of exchange energy. However, even though the intertwined wall is never a ground state, it is nevertheless a long-lived metastable state due to the pinning to the moir\'e lattice (see main text). 
\begin{figure}
	\includegraphics[trim={0cm 0cm 0cm 0cm}, width=0.6\linewidth,clip=true]{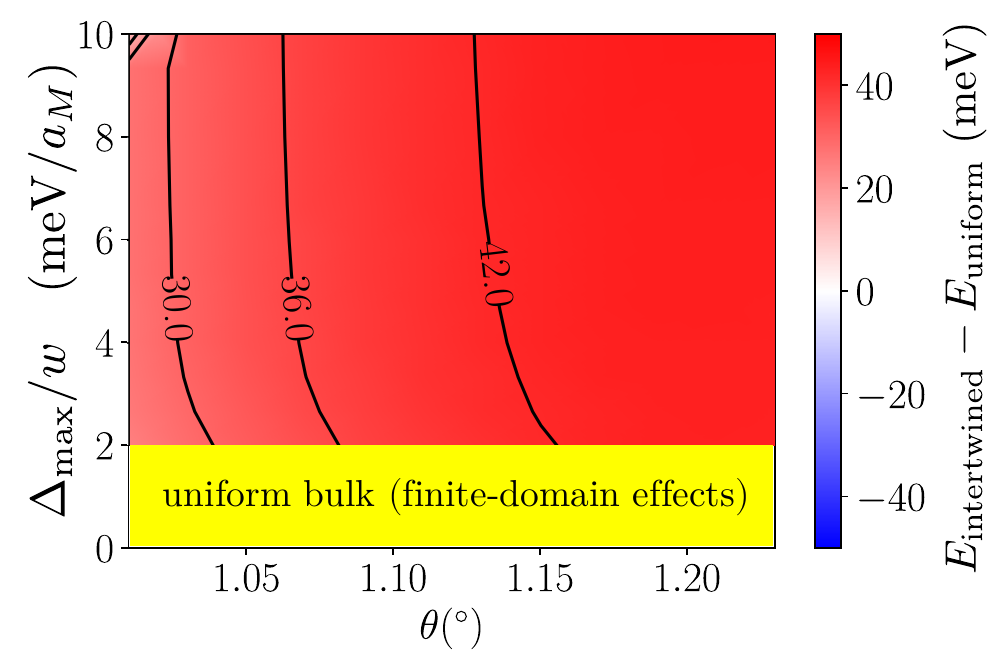}
	\caption{Phase diagram of the energy difference between intertwined wall and uniform solution, $E_\textrm{intertwined}-E_\textrm{uniform}$.  $N_1=N_2=10,2w=3a_\textrm{M}$.}
	\label{fig:phase_diagram_intertwined}
\end{figure}

\section{Local single-particle spectral function}\label{SecAppSpectral}
In this section we consider a DW solution $\Phi$, and compute the local single-particle spectral function (at $T=0$), which enters the expression for the tunneling conductance in a scanning tunneling microscopy experiment
\begin{equation}
    A_S(\omega,\bm{r})=\sum_{\gamma\tau s f}\bigg[
     \big|\bra{\gamma}\hat{\psi}^\dagger_{\tau s f}(\bm{r})\ket{\Phi}\big|^2\delta(\omega-E_\gamma+E_\Phi)+
     \big|\bra{\gamma}\hat{\psi}_{\tau s f}(\bm{r})\ket{\Phi}\big|^2\delta(\omega+E_\gamma-E_\Phi)
    \bigg]
\end{equation}
where $\gamma$ runs over excited states. We therefore require the following electron addition and removal matrix elements
\begin{equation}
    M_{e^-,\gamma}(\bm{r};\tau sf)\equiv \bra{\gamma}\hat{\psi}^\dagger_{\tau sf}(\bm{r})\ket{\Phi},\quad\quad
    M_{h^+,\gamma}(\bm{r};\tau sf)\equiv\bra{\gamma}\hat{\psi}_{\tau sf}(\bm{r})\ket{\Phi}.
\end{equation}
We approximate these quantities by their HF values. So $\ket{\Phi}$ is the (metastable) HF (DW) solution, and the relevant $\ket{\gamma}$ are the $N-1$ and $N+1$ particle states obtained by emptying a filled HF orbital or filling an empty HF orbital. The excited state energies are obtained via Koopman's theorem, e.g. assuming no rearrangement, for electron addition we have $E_\gamma-E_\Phi=\epsilon_\gamma$ where$\epsilon_\gamma$ is the HF eigenvalue .

We have three different bases in the setup with $k_2$ conservation. There is the real space basis $\hat{\psi}^\dagger_{\tau s f}(\bm{r})$, the bare basis $\hat{c}^\dagger_{\bm{k}\tau a s}$ obtained from the CM, and the HF basis $\hat{d}^\dagger_{k_2 n s}$ where the index $n$ can mix different $k_1,\tau,a$. We assume that the HF orbitals do not mix spins. The bases are related as follows
\begin{gather}
    \hat{c}^\dagger_{\bm{k}\tau a s}=\sum_f\int d\bm{r}\,\psi_{\bm{k}\tau as f}(\bm{r})\hat{\psi}^\dagger_{\tau s f}(\bm{r})\\
    \hat{\psi}^\dagger_{\tau sf}(\bm{r})=\sum_{\bm{k}a}\psi^*_{\bm{k}\tau a sf}(\bm{r})\hat{c}^\dagger_{\bm{k}\tau as}\\
    \hat{d}^\dagger_{k_2 ns}=\sum_{k_1\tau a}f_{n;k_1\tau a}(k_2;s)\hat{c}^\dagger_{\bm{k}\tau as}\\
    \hat{c}^\dagger_{\bm{k}\tau as}=\sum_n f^*_{n;k_1\tau a}(k_2;s)\hat{d}^\dagger_{k_2 ns}.
\end{gather}
where $\psi_{\bm{k}\tau asf}(\bm{r})$ are the zero-substrate CM wavefunctions with the layer/sublattice $f$ included as an index, and $f_{n;k_1\tau a}(k_2;s)$ parameterizes the HF orbitals.

The matrix elements are then (where $\gamma$ indicates the HF orbital in question, e.g. for electron addition $\ket{\gamma=(k_2ns)}=\hat{d}^\dagger_{k_2 ns}\ket{\Phi}$)
\begin{align}
    M_{e^-,(k_2 ns)}(\bm{r};\tau s f)&=\bra{\Phi}\hat{d}^{\phantom{\dagger}}_{k_2ns}\hat{\psi}^\dagger_{\tau sf}(\bm{r})\ket{\Phi}\\
    &=\sum_{k_1'\tau'a'}\sum_{k_1''a''}f^*_{n;k_1'\tau'a'}(k_2;s)\psi^*_{(k_1''k_2)\tau a'' sf}(\bm{r})\big[\delta_{k_1'k_1''}\delta_{\tau\tau'}\delta_{a'a''}-P_{k_1''\tau a''s;k_1'\tau' a's}(k_2)\big]\\
    M_{h^+,(k_2ns)}(\bm{r};\tau sf)&=\bra{\Phi}\hat{d}^\dagger_{k_2ns}\hat{\psi}^{\phantom{\dagger}}_{\tau sf}(\bm{r})\ket{\Phi}\\
    &=\sum_{k_1'\tau'a'}\sum_{k_1''a''}f_{n;k_1'\tau'a'}(k_2;s)\psi_{(k_1''k_2)\tau a'' sf}(\bm{r})P_{k_1'\tau'a's;k_1''\tau a''s}(k_2;s).
\end{align}

\section{Domain wall details}\label{SecAppDWDetails}

In Figs.~\ref{SMFigChernDW},\ref{SMFigThirdDW},\ref{SMFigValleyDW}, we show additional information on representative Chern, intertwined and valley DW solutions. The system size was chosen to be $N_1=N_2=20$ so that the chiral gapless modes can be more easily seen for the chiral DWs. If $N_2$ is too small, the momentum resolution is not fine enough to discern the chiral modes, and if $N_1$ is too small, the chiral modes can gap by hybridizing with their counterparts across the bulks. The substrate for the Chern and valley DWs was chosen to have sharp steps between $\pm 20\text{meV}$ at $x_1/a_\text{M}=0,10$, while the substrate is completely uniform for the intertwined wall. Note that the HFWQ overlaps have fast oscillations for the HFWQ basis corresponding to the `wrong' sign of sublattice mass---this is due to the fact that the HFWQ bases for $\pm \Delta_\text{max}$ are not generally orthogonal. The interaction potential used was the dual-gate screened form---for STM the single-gate screened interaction is more appropriate, but we have checked that the choice does not affect our results significantly. 
\begin{figure}
	\includegraphics[trim={0cm 0cm 0cm 0cm}, width=1\linewidth,clip=true]{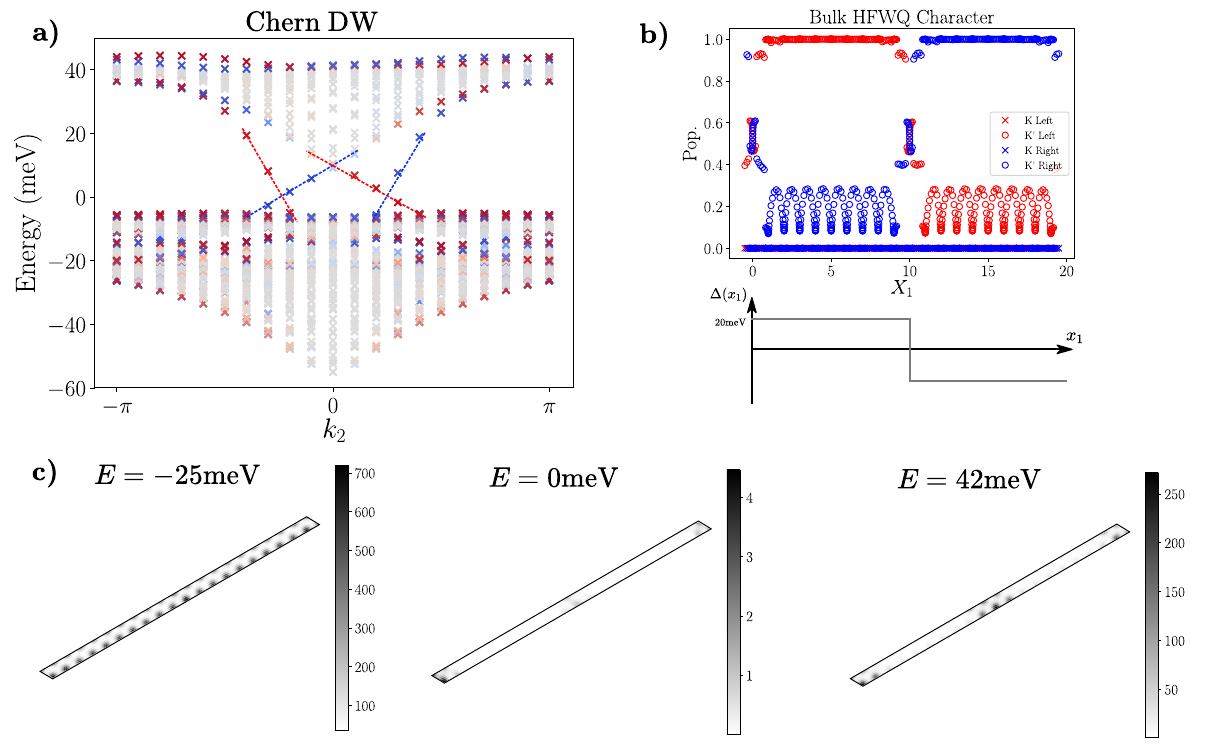}
	\caption{(a) HF bandstructure of the Chern DW organized by momentum parallel to the DW. The substrate $\Delta_2(\bm{r})$ is $\Delta_\text{max}=20$~meV in region $L$ and $-\Delta_\text{max}$ in region $R$. Points are colour-coded according to the real-space localization of the corresponding HF orbitals (cf. Fig.~2 in the main text). Dashed lines are guides indicating gapless chiral modes. (b) Overlap of the HFWQ orbitals (obtained using a uniform $\Delta=\Delta_\text{max}$ with the converged DW solution, plotted against the average $x_1$ position (in units of $a_\text{M}$) of the orbitals. $K\,\text{Left}$, $K\,\text{Right}$ refer to the $K$-polarized HFWQ basis constructed in a uniform substrate $\pm\Delta_\text{max}$ respectively, and analogously for valley $K'$. Note that the $L$ and $R$ bases for a given valley are not orthogonal, which explains the moir\'e scale oscillations. Bottom shows a schematic of the substrate profile. (c) Line cuts of the spatially-resolved spectral function (relevant for STM) at selected energies;  only one moir\'e cell in the $\mathbf{a}_\text{M}^{(2)}$-direction is shown as the results are periodic. Darker regions indicate higher weight. Parameters are $N_1=N_2=20,\theta=1.2^\circ$, and the dual-gate screened potential is used.\label{SMFigChernDW}}
\end{figure}

\begin{figure}
	\includegraphics[trim={0cm 0cm 0cm 0cm}, width=1\linewidth,clip=true]{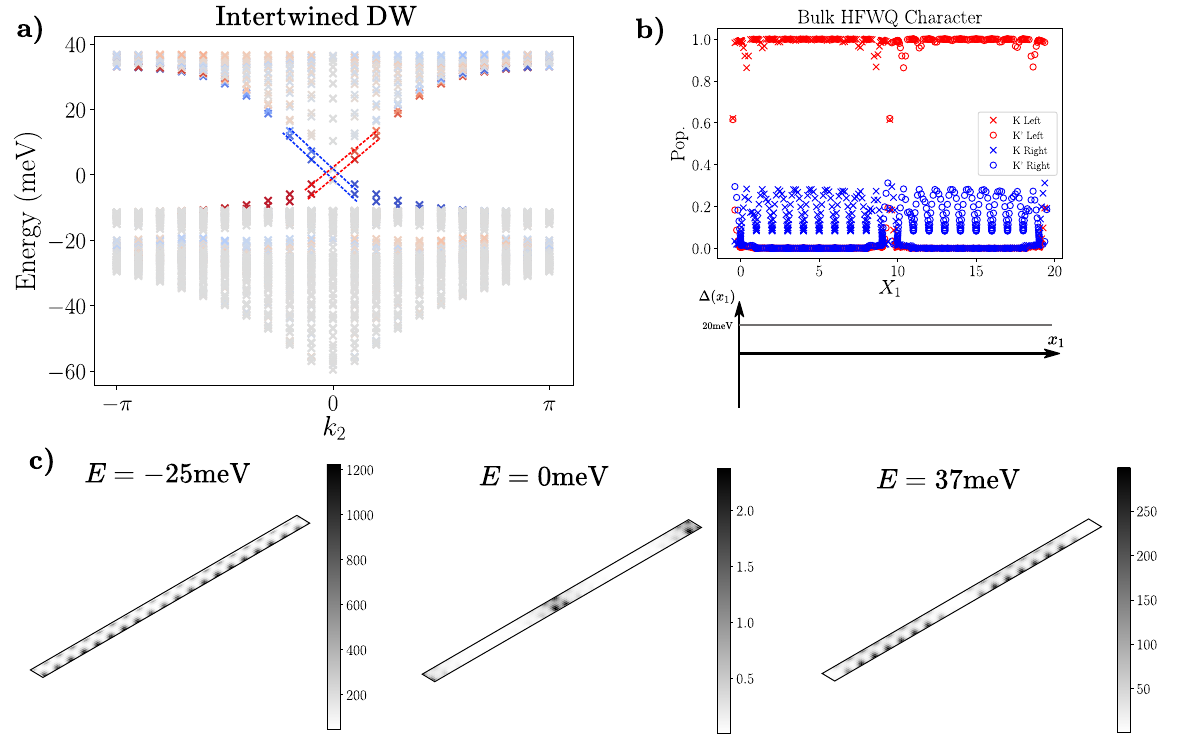}
	\caption{Same as Fig.~\ref{SMFigChernDW} except for the intertwined wall. It is evident from the HFWQ overlaps and the $E=0\text{meV}$ spatially-resolved density of states that the DW prefers to lock to half-integer positions.\label{SMFigThirdDW}}
\end{figure}

\begin{figure}
	\includegraphics[trim={0cm 0cm 0cm 0cm}, width=1\linewidth,clip=true]{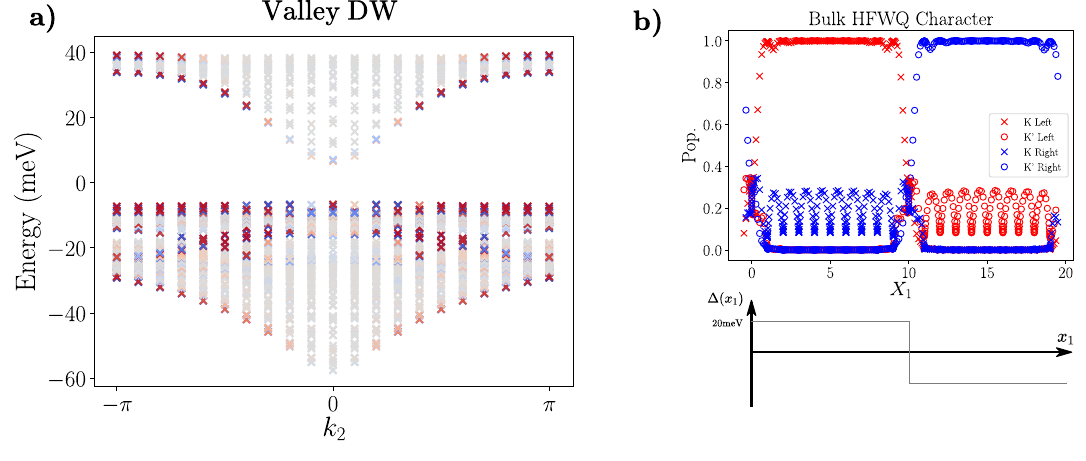}
	\caption{Same as Fig.~\ref{SMFigChernDW} except for the valley wall. At HF level the counter-propagating modes of the valley DW can hybridize and gap out. However a gapless mode must reemerge as quantum fluctuations restore $U(1)_v$ symmetry along the DW. Hence we have omitted plots of the spectral function.\label{SMFigValleyDW}}
\end{figure}

\section{Finite temperature phase transition}\label{SecAppChernValley}
We have determined the phase diagram for the competition between Chern and valley wall, and from the energy difference of the two solutions we can extract a domain wall tension $   \epsilon=(E_\textrm{Chern}-E_\textrm{valley})/(2a_MN_2)$. The factor of two arises because we have two DWs in our HF solution and $a_MN_2$ is the transverse length of the DW. For standard entropic reasons, we expect both walls to co-exist at high enough temperatures. Let us consider the ``substrate dominated" regime where the substrate is so large that the bulk states just track the local substrate. Further let us assume the substrate varies on a length scale $\xi_\textrm{dis}$. The energy cost of a Chern wall over a valley wall is then $\mu=\xi_\textrm{dis}\epsilon$. Here, we consider a simple model where the substrate has a checkerboard pattern with squares of side-length $\xi_\textrm{dis}\times\xi_\textrm{dis}$. Let us work with domain variables $\tau_i$ living on the plaquettes of the square lattice. If we assume that $\mu>0$ so that a valley DW network $|\Phi\rangle$ is the ground state then we can locally define $\tau_i=+1$ if the domain matches the corresponding domain in  $|\Phi\rangle$. So the ground state is all $\tau_i=1$ or all $\tau_i=-1$. If we have a mix of Chern and valley walls, then the Hamiltonian describing the energy cost is the 2d Ising model on a square lattice $
    H=-\mu\sum_{<i,j>}\tau_i\tau_j.
$
We have a finite-temperature phase transition at $
    T_c=2\mu/\log(1+\sqrt{2}).$ At $\theta=1.2^\circ$ and at small substrate slopes, we have a typical DW tension of $\epsilon\sim0.5$meV$/a_M$. This gives us a critical temperature $
    T_c\sim 13\textrm{K}\times\frac{\xi_\textrm{dis}}{a_M}.$
Given that the experiments are done at a temperature of $T\sim 2$K and the typical domain size is some mesoscopic scale $\xi_\textrm{dis}\sim1\mu$m, this phase transition occurs at too large temperatures to be relevant. So unless we have some fine-tuning such that we are close to the phase boundary in Fig.1 of the main paper, we will generically have either only valley or only Chern walls in our system.

\clearpage
}

\twocolumngrid

	\newpage

	\newpage

\clearpage


\begin{thebibliography}{50}%
\makeatletter
\providecommand \@ifxundefined [1]{%
 \@ifx{#1\undefined}
}%
\providecommand \@ifnum [1]{%
 \ifnum #1\expandafter \@firstoftwo
 \else \expandafter \@secondoftwo
 \fi
}%
\providecommand \@ifx [1]{%
 \ifx #1\expandafter \@firstoftwo
 \else \expandafter \@secondoftwo
 \fi
}%
\providecommand \natexlab [1]{#1}%
\providecommand \enquote  [1]{``#1''}%
\providecommand \bibnamefont  [1]{#1}%
\providecommand \bibfnamefont [1]{#1}%
\providecommand \citenamefont [1]{#1}%
\providecommand \href@noop [0]{\@secondoftwo}%
\providecommand \href [0]{\begingroup \@sanitize@url \@href}%
\providecommand \@href[1]{\@@startlink{#1}\@@href}%
\providecommand \@@href[1]{\endgroup#1\@@endlink}%
\providecommand \@sanitize@url [0]{\catcode `\\12\catcode `\$12\catcode
  `\&12\catcode `\#12\catcode `\^12\catcode `\_12\catcode `\%12\relax}%
\providecommand \@@startlink[1]{}%
\providecommand \@@endlink[0]{}%
\providecommand \url  [0]{\begingroup\@sanitize@url \@url }%
\providecommand \@url [1]{\endgroup\@href {#1}{\urlprefix }}%
\providecommand \urlprefix  [0]{URL }%
\providecommand \Eprint [0]{\href }%
\providecommand \doibase [0]{https://doi.org/}%
\providecommand \selectlanguage [0]{\@gobble}%
\providecommand \bibinfo  [0]{\@secondoftwo}%
\providecommand \bibfield  [0]{\@secondoftwo}%
\providecommand \translation [1]{[#1]}%
\providecommand \BibitemOpen [0]{}%
\providecommand \bibitemStop [0]{}%
\providecommand \bibitemNoStop [0]{.\EOS\space}%
\providecommand \EOS [0]{\spacefactor3000\relax}%
\providecommand \BibitemShut  [1]{\csname bibitem#1\endcsname}%
\let\auto@bib@innerbib\@empty
\bibitem [{\citenamefont {Cao}\ \emph {et~al.}(2018{\natexlab{a}})\citenamefont
  {Cao}, \citenamefont {Fatemi}, \citenamefont {Demir}, \citenamefont {Fang},
  \citenamefont {Tomarken}, \citenamefont {Luo}, \citenamefont
  {Sanchez-Yamagishi}, \citenamefont {Watanabe}, \citenamefont {Taniguchi},
  \citenamefont {Kaxiras}, \citenamefont {Ashoori},\ and\ \citenamefont
  {Jarillo-Herrero}}]{Cao2018}%
  \BibitemOpen
  \bibfield  {author} {\bibinfo {author} {\bibfnamefont {Y.}~\bibnamefont
  {Cao}}, \bibinfo {author} {\bibfnamefont {V.}~\bibnamefont {Fatemi}},
  \bibinfo {author} {\bibfnamefont {A.}~\bibnamefont {Demir}}, \bibinfo
  {author} {\bibfnamefont {S.}~\bibnamefont {Fang}}, \bibinfo {author}
  {\bibfnamefont {S.~L.}\ \bibnamefont {Tomarken}}, \bibinfo {author}
  {\bibfnamefont {J.~Y.}\ \bibnamefont {Luo}}, \bibinfo {author} {\bibfnamefont
  {J.~D.}\ \bibnamefont {Sanchez-Yamagishi}}, \bibinfo {author} {\bibfnamefont
  {K.}~\bibnamefont {Watanabe}}, \bibinfo {author} {\bibfnamefont
  {T.}~\bibnamefont {Taniguchi}}, \bibinfo {author} {\bibfnamefont
  {E.}~\bibnamefont {Kaxiras}}, \bibinfo {author} {\bibfnamefont {R.~C.}\
  \bibnamefont {Ashoori}},\ and\ \bibinfo {author} {\bibfnamefont
  {P.}~\bibnamefont {Jarillo-Herrero}},\ }\bibfield  {title} {\bibinfo {title}
  {Correlated insulator behaviour at half-filling in magic-angle graphene
  superlattices},\ }\href {https://doi.org/10.1038/nature26154} {\bibfield
  {journal} {\bibinfo  {journal} {Nature}\ }\textbf {\bibinfo {volume} {556}},\
  \bibinfo {pages} {80} (\bibinfo {year} {2018}{\natexlab{a}})}\BibitemShut
  {NoStop}%
\bibitem [{\citenamefont {Cao}\ \emph {et~al.}(2018{\natexlab{b}})\citenamefont
  {Cao}, \citenamefont {Fatemi}, \citenamefont {Fang}, \citenamefont
  {Watanabe}, \citenamefont {Taniguchi}, \citenamefont {Kaxiras},\ and\
  \citenamefont {Jarillo-Herrero}}]{Cao2018b}%
  \BibitemOpen
  \bibfield  {author} {\bibinfo {author} {\bibfnamefont {Y.}~\bibnamefont
  {Cao}}, \bibinfo {author} {\bibfnamefont {V.}~\bibnamefont {Fatemi}},
  \bibinfo {author} {\bibfnamefont {S.}~\bibnamefont {Fang}}, \bibinfo {author}
  {\bibfnamefont {K.}~\bibnamefont {Watanabe}}, \bibinfo {author}
  {\bibfnamefont {T.}~\bibnamefont {Taniguchi}}, \bibinfo {author}
  {\bibfnamefont {E.}~\bibnamefont {Kaxiras}},\ and\ \bibinfo {author}
  {\bibfnamefont {P.}~\bibnamefont {Jarillo-Herrero}},\ }\bibfield  {title}
  {\bibinfo {title} {Unconventional superconductivity in magic-angle graphene
  superlattices},\ }\href {https://doi.org/10.1038/nature26160} {\bibfield
  {journal} {\bibinfo  {journal} {Nature}\ }\textbf {\bibinfo {volume} {556}},\
  \bibinfo {pages} {43} (\bibinfo {year} {2018}{\natexlab{b}})}\BibitemShut
  {NoStop}%
\bibitem [{\citenamefont {Yankowitz}\ \emph {et~al.}(2019)\citenamefont
  {Yankowitz}, \citenamefont {Chen}, \citenamefont {Polshyn}, \citenamefont
  {Zhang}, \citenamefont {Watanabe}, \citenamefont {Taniguchi}, \citenamefont
  {Graf}, \citenamefont {Young},\ and\ \citenamefont {Dean}}]{Yankowitz1059}%
  \BibitemOpen
  \bibfield  {author} {\bibinfo {author} {\bibfnamefont {M.}~\bibnamefont
  {Yankowitz}}, \bibinfo {author} {\bibfnamefont {S.}~\bibnamefont {Chen}},
  \bibinfo {author} {\bibfnamefont {H.}~\bibnamefont {Polshyn}}, \bibinfo
  {author} {\bibfnamefont {Y.}~\bibnamefont {Zhang}}, \bibinfo {author}
  {\bibfnamefont {K.}~\bibnamefont {Watanabe}}, \bibinfo {author}
  {\bibfnamefont {T.}~\bibnamefont {Taniguchi}}, \bibinfo {author}
  {\bibfnamefont {D.}~\bibnamefont {Graf}}, \bibinfo {author} {\bibfnamefont
  {A.~F.}\ \bibnamefont {Young}},\ and\ \bibinfo {author} {\bibfnamefont
  {C.~R.}\ \bibnamefont {Dean}},\ }\bibfield  {title} {\bibinfo {title} {Tuning
  superconductivity in twisted bilayer graphene},\ }\href
  {https://doi.org/10.1126/science.aav1910} {\bibfield  {journal} {\bibinfo
  {journal} {Science}\ }\textbf {\bibinfo {volume} {363}},\ \bibinfo {pages}
  {1059} (\bibinfo {year} {2019})}\BibitemShut {NoStop}%
\bibitem [{\citenamefont {Sharpe}\ \emph {et~al.}(2019)\citenamefont {Sharpe},
  \citenamefont {Fox}, \citenamefont {Barnard}, \citenamefont {Finney},
  \citenamefont {Watanabe}, \citenamefont {Taniguchi}, \citenamefont
  {Kastner},\ and\ \citenamefont {Goldhaber-Gordon}}]{Sharpe605}%
  \BibitemOpen
  \bibfield  {author} {\bibinfo {author} {\bibfnamefont {A.~L.}\ \bibnamefont
  {Sharpe}}, \bibinfo {author} {\bibfnamefont {E.~J.}\ \bibnamefont {Fox}},
  \bibinfo {author} {\bibfnamefont {A.~W.}\ \bibnamefont {Barnard}}, \bibinfo
  {author} {\bibfnamefont {J.}~\bibnamefont {Finney}}, \bibinfo {author}
  {\bibfnamefont {K.}~\bibnamefont {Watanabe}}, \bibinfo {author}
  {\bibfnamefont {T.}~\bibnamefont {Taniguchi}}, \bibinfo {author}
  {\bibfnamefont {M.~A.}\ \bibnamefont {Kastner}},\ and\ \bibinfo {author}
  {\bibfnamefont {D.}~\bibnamefont {Goldhaber-Gordon}},\ }\bibfield  {title}
  {\bibinfo {title} {Emergent ferromagnetism near three-quarters filling in
  twisted bilayer graphene},\ }\href {https://doi.org/10.1126/science.aaw3780}
  {\bibfield  {journal} {\bibinfo  {journal} {Science}\ }\textbf {\bibinfo
  {volume} {365}},\ \bibinfo {pages} {605} (\bibinfo {year}
  {2019})}\BibitemShut {NoStop}%
\bibitem [{\citenamefont {Serlin}\ \emph {et~al.}(2020)\citenamefont {Serlin},
  \citenamefont {Tschirhart}, \citenamefont {Polshyn}, \citenamefont {Zhang},
  \citenamefont {Zhu}, \citenamefont {Watanabe}, \citenamefont {Taniguchi},
  \citenamefont {Balents},\ and\ \citenamefont {Young}}]{Serlin}%
  \BibitemOpen
  \bibfield  {author} {\bibinfo {author} {\bibfnamefont {M.}~\bibnamefont
  {Serlin}}, \bibinfo {author} {\bibfnamefont {C.~L.}\ \bibnamefont
  {Tschirhart}}, \bibinfo {author} {\bibfnamefont {H.}~\bibnamefont {Polshyn}},
  \bibinfo {author} {\bibfnamefont {Y.}~\bibnamefont {Zhang}}, \bibinfo
  {author} {\bibfnamefont {J.}~\bibnamefont {Zhu}}, \bibinfo {author}
  {\bibfnamefont {K.}~\bibnamefont {Watanabe}}, \bibinfo {author}
  {\bibfnamefont {T.}~\bibnamefont {Taniguchi}}, \bibinfo {author}
  {\bibfnamefont {L.}~\bibnamefont {Balents}},\ and\ \bibinfo {author}
  {\bibfnamefont {A.~F.}\ \bibnamefont {Young}},\ }\bibfield  {title} {\bibinfo
  {title} {Intrinsic quantized anomalous hall effect in a moir{\'e}
  heterostructure},\ }\href
  {https://science.sciencemag.org/content/early/2019/12/20/science.aay5533}
  {\bibfield  {journal} {\bibinfo  {journal} {Science}\ }\textbf {\bibinfo
  {volume} {367}},\ \bibinfo {pages} {900} (\bibinfo {year}
  {2020})}\BibitemShut {NoStop}%
\bibitem [{\citenamefont {Tschirhart}\ \emph {et~al.}(2020)\citenamefont
  {Tschirhart}, \citenamefont {Serlin}, \citenamefont {Polshyn}, \citenamefont
  {Shragai}, \citenamefont {Xia}, \citenamefont {Zhu}, \citenamefont {Zhang},
  \citenamefont {Watanabe}, \citenamefont {Taniguchi}, \citenamefont {Huber},\
  and\ \citenamefont {Young}}]{tschirhart2020imaging}%
  \BibitemOpen
  \bibfield  {author} {\bibinfo {author} {\bibfnamefont {C.~L.}\ \bibnamefont
  {Tschirhart}}, \bibinfo {author} {\bibfnamefont {M.}~\bibnamefont {Serlin}},
  \bibinfo {author} {\bibfnamefont {H.}~\bibnamefont {Polshyn}}, \bibinfo
  {author} {\bibfnamefont {A.}~\bibnamefont {Shragai}}, \bibinfo {author}
  {\bibfnamefont {Z.}~\bibnamefont {Xia}}, \bibinfo {author} {\bibfnamefont
  {J.}~\bibnamefont {Zhu}}, \bibinfo {author} {\bibfnamefont {Y.}~\bibnamefont
  {Zhang}}, \bibinfo {author} {\bibfnamefont {K.}~\bibnamefont {Watanabe}},
  \bibinfo {author} {\bibfnamefont {T.}~\bibnamefont {Taniguchi}}, \bibinfo
  {author} {\bibfnamefont {M.~E.}\ \bibnamefont {Huber}},\ and\ \bibinfo
  {author} {\bibfnamefont {A.~F.}\ \bibnamefont {Young}},\ }\href@noop {}
  {\bibinfo {title} {Imaging orbital ferromagnetism in a moiré chern
  insulator}} (\bibinfo {year} {2020}),\ \Eprint
  {https://arxiv.org/abs/2006.08053} {arXiv:2006.08053 [cond-mat.mes-hall]}
  \BibitemShut {NoStop}%
\bibitem [{\citenamefont {Bultinck}\ \emph {et~al.}(2020)\citenamefont
  {Bultinck}, \citenamefont {Chatterjee},\ and\ \citenamefont
  {Zaletel}}]{bultinck2020mechanism}%
  \BibitemOpen
  \bibfield  {author} {\bibinfo {author} {\bibfnamefont {N.}~\bibnamefont
  {Bultinck}}, \bibinfo {author} {\bibfnamefont {S.}~\bibnamefont
  {Chatterjee}},\ and\ \bibinfo {author} {\bibfnamefont {M.~P.}\ \bibnamefont
  {Zaletel}},\ }\bibfield  {title} {\bibinfo {title} {Mechanism for anomalous
  hall ferromagnetism in twisted bilayer graphene},\ }\href
  {https://journals.aps.org/prl/abstract/10.1103/PhysRevLett.124.166601}
  {\bibfield  {journal} {\bibinfo  {journal} {Physical Review Letters}\
  }\textbf {\bibinfo {volume} {124}},\ \bibinfo {pages} {166601} (\bibinfo
  {year} {2020})}\BibitemShut {NoStop}%
\bibitem [{\citenamefont {Liu}\ and\ \citenamefont {Dai}(2019)}]{Xi2019}%
  \BibitemOpen
  \bibfield  {author} {\bibinfo {author} {\bibfnamefont {J.}~\bibnamefont
  {Liu}}\ and\ \bibinfo {author} {\bibfnamefont {X.}~\bibnamefont {Dai}},\
  }\bibfield  {title} {\bibinfo {title} {Correlated insulating states and the
  quantum anomalous hall phenomena at all integer fillings in twisted bilayer
  graphene},\ }\href {https://arxiv.org/abs/1911.03760} {\bibfield  {journal}
  {\bibinfo  {journal} {arXiv preprint arXiv:1911.03760}\ } (\bibinfo {year}
  {2019})}\BibitemShut {NoStop}%
\bibitem [{\citenamefont {Zhang}\ \emph {et~al.}(2019)\citenamefont {Zhang},
  \citenamefont {Mao},\ and\ \citenamefont {Senthil}}]{zhang2019twisted}%
  \BibitemOpen
  \bibfield  {author} {\bibinfo {author} {\bibfnamefont {Y.-H.}\ \bibnamefont
  {Zhang}}, \bibinfo {author} {\bibfnamefont {D.}~\bibnamefont {Mao}},\ and\
  \bibinfo {author} {\bibfnamefont {T.}~\bibnamefont {Senthil}},\ }\bibfield
  {title} {\bibinfo {title} {Twisted bilayer graphene aligned with hexagonal
  boron nitride: anomalous hall effect and a lattice model},\ }\href@noop {}
  {\bibfield  {journal} {\bibinfo  {journal} {Physical Review Research}\
  }\textbf {\bibinfo {volume} {1}},\ \bibinfo {pages} {033126} (\bibinfo {year}
  {2019})}\BibitemShut {NoStop}%
\bibitem [{\citenamefont {Bistritzer}\ and\ \citenamefont
  {MacDonald}(2011)}]{Bistritzer}%
  \BibitemOpen
  \bibfield  {author} {\bibinfo {author} {\bibfnamefont {R.}~\bibnamefont
  {Bistritzer}}\ and\ \bibinfo {author} {\bibfnamefont {A.~H.}\ \bibnamefont
  {MacDonald}},\ }\bibfield  {title} {\bibinfo {title} {Moir{\'e} bands in
  twisted double-layer graphene},\ }\href
  {https://doi.org/10.1073/pnas.1108174108} {\bibfield  {journal} {\bibinfo
  {journal} {Proceedings of the National Academy of Sciences}\ }\textbf
  {\bibinfo {volume} {108}},\ \bibinfo {pages} {12233} (\bibinfo {year}
  {2011})}\BibitemShut {NoStop}%
\bibitem [{\citenamefont {Jung}\ \emph {et~al.}(2015)\citenamefont {Jung},
  \citenamefont {DaSilva}, \citenamefont {MacDonald},\ and\ \citenamefont
  {Adam}}]{jung2015origin}%
  \BibitemOpen
  \bibfield  {author} {\bibinfo {author} {\bibfnamefont {J.}~\bibnamefont
  {Jung}}, \bibinfo {author} {\bibfnamefont {A.~M.}\ \bibnamefont {DaSilva}},
  \bibinfo {author} {\bibfnamefont {A.~H.}\ \bibnamefont {MacDonald}},\ and\
  \bibinfo {author} {\bibfnamefont {S.}~\bibnamefont {Adam}},\ }\bibfield
  {title} {\bibinfo {title} {Origin of band gaps in graphene on hexagonal boron
  nitride},\ }\href {https://www.nature.com/articles/ncomms7308} {\bibfield
  {journal} {\bibinfo  {journal} {Nature communications}\ }\textbf {\bibinfo
  {volume} {6}},\ \bibinfo {pages} {1} (\bibinfo {year} {2015})}\BibitemShut
  {NoStop}%
\bibitem [{\citenamefont {Kumar}\ \emph {et~al.}(2014)\citenamefont {Kumar},
  \citenamefont {Roy},\ and\ \citenamefont {Sondhi}}]{QHFMChern}%
  \BibitemOpen
  \bibfield  {author} {\bibinfo {author} {\bibfnamefont {A.}~\bibnamefont
  {Kumar}}, \bibinfo {author} {\bibfnamefont {R.}~\bibnamefont {Roy}},\ and\
  \bibinfo {author} {\bibfnamefont {S.~L.}\ \bibnamefont {Sondhi}},\ }\bibfield
   {title} {\bibinfo {title} {Generalizing quantum hall ferromagnetism to
  fractional chern bands},\ }\href {https://doi.org/10.1103/PhysRevB.90.245106}
  {\bibfield  {journal} {\bibinfo  {journal} {Phys. Rev. B}\ }\textbf {\bibinfo
  {volume} {90}},\ \bibinfo {pages} {245106} (\bibinfo {year}
  {2014})}\BibitemShut {NoStop}%
\bibitem [{\citenamefont {Liu}\ \emph {et~al.}(2019{\natexlab{a}})\citenamefont
  {Liu}, \citenamefont {Khalaf}, \citenamefont {Lee},\ and\ \citenamefont
  {Vishwanath}}]{LiuHF2019}%
  \BibitemOpen
  \bibfield  {author} {\bibinfo {author} {\bibfnamefont {S.}~\bibnamefont
  {Liu}}, \bibinfo {author} {\bibfnamefont {E.}~\bibnamefont {Khalaf}},
  \bibinfo {author} {\bibfnamefont {J.~Y.}\ \bibnamefont {Lee}},\ and\ \bibinfo
  {author} {\bibfnamefont {A.}~\bibnamefont {Vishwanath}},\ }\bibfield  {title}
  {\bibinfo {title} {Nematic topological semimetal and insulator in magic angle
  bilayer graphene at charge neutrality},\ }\href
  {https://arxiv.org/abs/1905.07409} {\bibfield  {journal} {\bibinfo  {journal}
  {arXiv preprint arXiv:1905.07409}\ } (\bibinfo {year}
  {2019}{\natexlab{a}})}\BibitemShut {NoStop}%
\bibitem [{\citenamefont {Bultinck}\ \emph {et~al.}(2019)\citenamefont
  {Bultinck}, \citenamefont {Khalaf}, \citenamefont {Liu}, \citenamefont
  {Chatterjee}, \citenamefont {Vishwanath},\ and\ \citenamefont
  {Zaletel}}]{bultinck2_2019}%
  \BibitemOpen
  \bibfield  {author} {\bibinfo {author} {\bibfnamefont {N.}~\bibnamefont
  {Bultinck}}, \bibinfo {author} {\bibfnamefont {E.}~\bibnamefont {Khalaf}},
  \bibinfo {author} {\bibfnamefont {S.}~\bibnamefont {Liu}}, \bibinfo {author}
  {\bibfnamefont {S.}~\bibnamefont {Chatterjee}}, \bibinfo {author}
  {\bibfnamefont {A.}~\bibnamefont {Vishwanath}},\ and\ \bibinfo {author}
  {\bibfnamefont {M.~P.}\ \bibnamefont {Zaletel}},\ }\bibfield  {title}
  {\bibinfo {title} {Ground state and hidden symmetry of magic angle graphene
  at even integer filling},\ }\href {https://arxiv.org/abs/1911.02045}
  {\bibfield  {journal} {\bibinfo  {journal} {arXiv preprint arXiv:1911.02045}\
  } (\bibinfo {year} {2019})}\BibitemShut {NoStop}%
\bibitem [{\citenamefont {Zhang}\ \emph {et~al.}(2020)\citenamefont {Zhang},
  \citenamefont {Jiang}, \citenamefont {Wang},\ and\ \citenamefont
  {Zhang}}]{Yi2020}%
  \BibitemOpen
  \bibfield  {author} {\bibinfo {author} {\bibfnamefont {Y.}~\bibnamefont
  {Zhang}}, \bibinfo {author} {\bibfnamefont {K.}~\bibnamefont {Jiang}},
  \bibinfo {author} {\bibfnamefont {Z.}~\bibnamefont {Wang}},\ and\ \bibinfo
  {author} {\bibfnamefont {F.}~\bibnamefont {Zhang}},\ }\bibfield  {title}
  {\bibinfo {title} {Spontaneous symmetry breaking and topology in twisted
  bilayer graphene: the nature of the correlated insulating states and the
  quantum anomalous hall effect},\ }\href {https://arxiv.org/abs/2001.02476}
  {\bibfield  {journal} {\bibinfo  {journal} {arXiv preprint arXiv:2001.02476}\
  } (\bibinfo {year} {2020})}\BibitemShut {NoStop}%
\bibitem [{\citenamefont {Lin}\ and\ \citenamefont
  {Nandkishore}(2019)}]{lin2019}%
  \BibitemOpen
  \bibfield  {author} {\bibinfo {author} {\bibfnamefont {Y.-P.}\ \bibnamefont
  {Lin}}\ and\ \bibinfo {author} {\bibfnamefont {R.~M.}\ \bibnamefont
  {Nandkishore}},\ }\bibfield  {title} {\bibinfo {title} {Chiral twist on the
  high-${T}_{c}$ phase diagram in moir\'e heterostructures},\ }\href
  {https://doi.org/10.1103/PhysRevB.100.085136} {\bibfield  {journal} {\bibinfo
   {journal} {Phys. Rev. B}\ }\textbf {\bibinfo {volume} {100}},\ \bibinfo
  {pages} {085136} (\bibinfo {year} {2019})}\BibitemShut {NoStop}%
\bibitem [{\citenamefont {Xie}\ and\ \citenamefont {MacDonald}(2020)}]{Xie}%
  \BibitemOpen
  \bibfield  {author} {\bibinfo {author} {\bibfnamefont {M.}~\bibnamefont
  {Xie}}\ and\ \bibinfo {author} {\bibfnamefont {A.~H.}\ \bibnamefont
  {MacDonald}},\ }\bibfield  {title} {\bibinfo {title} {Nature of the
  correlated insulator states in twisted bilayer graphene},\ }\href
  {https://doi.org/10.1103/PhysRevLett.124.097601} {\bibfield  {journal}
  {\bibinfo  {journal} {Phys. Rev. Lett.}\ }\textbf {\bibinfo {volume} {124}},\
  \bibinfo {pages} {097601} (\bibinfo {year} {2020})}\BibitemShut {NoStop}%
\bibitem [{\citenamefont {Cea}\ and\ \citenamefont {Guinea}(2020)}]{Cea}%
  \BibitemOpen
  \bibfield  {author} {\bibinfo {author} {\bibfnamefont {T.}~\bibnamefont
  {Cea}}\ and\ \bibinfo {author} {\bibfnamefont {F.}~\bibnamefont {Guinea}},\
  }\bibfield  {title} {\bibinfo {title} {Band structure and insulating states
  driven by coulomb interaction in twisted bilayer graphene},\ }\href
  {https://doi.org/10.1103/PhysRevB.102.045107} {\bibfield  {journal} {\bibinfo
   {journal} {Phys. Rev. B}\ }\textbf {\bibinfo {volume} {102}},\ \bibinfo
  {pages} {045107} (\bibinfo {year} {2020})}\BibitemShut {NoStop}%
\bibitem [{\citenamefont {Lin}\ and\ \citenamefont {Ni}(2020)}]{Lin}%
  \BibitemOpen
  \bibfield  {author} {\bibinfo {author} {\bibfnamefont {X.}~\bibnamefont
  {Lin}}\ and\ \bibinfo {author} {\bibfnamefont {J.}~\bibnamefont {Ni}},\
  }\bibfield  {title} {\bibinfo {title} {Symmetry breaking in the double
  moir\'e superlattices of relaxed twisted bilayer graphene on hexagonal boron
  nitride},\ }\href {https://doi.org/10.1103/PhysRevB.102.035441} {\bibfield
  {journal} {\bibinfo  {journal} {Phys. Rev. B}\ }\textbf {\bibinfo {volume}
  {102}},\ \bibinfo {pages} {035441} (\bibinfo {year} {2020})}\BibitemShut
  {NoStop}%
\bibitem [{\citenamefont {Kwan}\ \emph {et~al.}(2020)\citenamefont {Kwan},
  \citenamefont {Hu}, \citenamefont {Simon},\ and\ \citenamefont
  {Parameswaran}}]{kwan2020excitonic}%
  \BibitemOpen
  \bibfield  {author} {\bibinfo {author} {\bibfnamefont {Y.~H.}\ \bibnamefont
  {Kwan}}, \bibinfo {author} {\bibfnamefont {Y.}~\bibnamefont {Hu}}, \bibinfo
  {author} {\bibfnamefont {S.~H.}\ \bibnamefont {Simon}},\ and\ \bibinfo
  {author} {\bibfnamefont {S.~A.}\ \bibnamefont {Parameswaran}},\ }\href@noop
  {} {\bibinfo {title} {Excitonic fractional quantum hall hierarchy in moiré
  heterostructures}} (\bibinfo {year} {2020}),\ \Eprint
  {https://arxiv.org/abs/2003.11559} {arXiv:2003.11559 [cond-mat.str-el]}
  \BibitemShut {NoStop}%
\bibitem [{\citenamefont {Stefanidis}\ and\ \citenamefont
  {Sodemann}(2020)}]{stefanidis2020excitonic}%
  \BibitemOpen
  \bibfield  {author} {\bibinfo {author} {\bibfnamefont {N.}~\bibnamefont
  {Stefanidis}}\ and\ \bibinfo {author} {\bibfnamefont {I.}~\bibnamefont
  {Sodemann}},\ }\href@noop {} {\bibinfo {title} {Excitonic laughlin states in
  ideal topological insulator flat bands and possible presence in moiré
  superlattice materials}} (\bibinfo {year} {2020}),\ \Eprint
  {https://arxiv.org/abs/2004.03613} {arXiv:2004.03613 [cond-mat.str-el]}
  \BibitemShut {NoStop}%
\bibitem [{\citenamefont {Cea}\ \emph {et~al.}(2020)\citenamefont {Cea},
  \citenamefont {Pantaleon},\ and\ \citenamefont {Guinea}}]{cea2020hBN}%
  \BibitemOpen
  \bibfield  {author} {\bibinfo {author} {\bibfnamefont {T.}~\bibnamefont
  {Cea}}, \bibinfo {author} {\bibfnamefont {P.~A.}\ \bibnamefont {Pantaleon}},\
  and\ \bibinfo {author} {\bibfnamefont {F.}~\bibnamefont {Guinea}},\
  }\href@noop {} {\bibinfo {title} {Band structure of twisted bilayer graphene
  on hexagonal boron nitride}} (\bibinfo {year} {2020}),\ \Eprint
  {https://arxiv.org/abs/2005.07396} {arXiv:2005.07396 [cond-mat.str-el]}
  \BibitemShut {NoStop}%
\bibitem [{\citenamefont {Shi}\ \emph {et~al.}(2021)\citenamefont {Shi},
  \citenamefont {Zhu},\ and\ \citenamefont {MacDonald}}]{macdonald2021}%
  \BibitemOpen
  \bibfield  {author} {\bibinfo {author} {\bibfnamefont {J.}~\bibnamefont
  {Shi}}, \bibinfo {author} {\bibfnamefont {J.}~\bibnamefont {Zhu}},\ and\
  \bibinfo {author} {\bibfnamefont {A.~H.}\ \bibnamefont {MacDonald}},\
  }\bibfield  {title} {\bibinfo {title} {Moir\'e commensurability and the
  quantum anomalous hall effect in twisted bilayer graphene on hexagonal boron
  nitride},\ }\href {https://doi.org/10.1103/PhysRevB.103.075122} {\bibfield
  {journal} {\bibinfo  {journal} {Phys. Rev. B}\ }\textbf {\bibinfo {volume}
  {103}},\ \bibinfo {pages} {075122} (\bibinfo {year} {2021})}\BibitemShut
  {NoStop}%
\bibitem [{\citenamefont {Huang}\ \emph {et~al.}(2018)\citenamefont {Huang},
  \citenamefont {Kim}, \citenamefont {Efimkin}, \citenamefont {Lovorn},
  \citenamefont {Taniguchi}, \citenamefont {Watanabe}, \citenamefont
  {MacDonald}, \citenamefont {Tutuc},\ and\ \citenamefont {LeRoy}}]{huang2018}%
  \BibitemOpen
  \bibfield  {author} {\bibinfo {author} {\bibfnamefont {S.}~\bibnamefont
  {Huang}}, \bibinfo {author} {\bibfnamefont {K.}~\bibnamefont {Kim}}, \bibinfo
  {author} {\bibfnamefont {D.~K.}\ \bibnamefont {Efimkin}}, \bibinfo {author}
  {\bibfnamefont {T.}~\bibnamefont {Lovorn}}, \bibinfo {author} {\bibfnamefont
  {T.}~\bibnamefont {Taniguchi}}, \bibinfo {author} {\bibfnamefont
  {K.}~\bibnamefont {Watanabe}}, \bibinfo {author} {\bibfnamefont {A.~H.}\
  \bibnamefont {MacDonald}}, \bibinfo {author} {\bibfnamefont {E.}~\bibnamefont
  {Tutuc}},\ and\ \bibinfo {author} {\bibfnamefont {B.~J.}\ \bibnamefont
  {LeRoy}},\ }\bibfield  {title} {\bibinfo {title} {Topologically protected
  helical states in minimally twisted bilayer graphene},\ }\href
  {https://doi.org/10.1103/PhysRevLett.121.037702} {\bibfield  {journal}
  {\bibinfo  {journal} {Phys. Rev. Lett.}\ }\textbf {\bibinfo {volume} {121}},\
  \bibinfo {pages} {037702} (\bibinfo {year} {2018})}\BibitemShut {NoStop}%
\bibitem [{\citenamefont {Fal'ko}\ and\ \citenamefont
  {Iordanskii}(1999)}]{Falko:1999p1}%
  \BibitemOpen
  \bibfield  {author} {\bibinfo {author} {\bibfnamefont {V.~I.}\ \bibnamefont
  {Fal'ko}}\ and\ \bibinfo {author} {\bibfnamefont {S.~V.}\ \bibnamefont
  {Iordanskii}},\ }\bibfield  {title} {\bibinfo {title} {Topological defects
  and goldstone excitations in domain walls between ferromagnetic quantum hall
  liquids},\ }\href {https://doi.org/10.1103/PhysRevLett.82.402} {\bibfield
  {journal} {\bibinfo  {journal} {Phys. Rev. Lett.}\ }\textbf {\bibinfo
  {volume} {82}},\ \bibinfo {pages} {402} (\bibinfo {year} {1999})}\BibitemShut
  {NoStop}%
\bibitem [{\citenamefont {Mitra}\ and\ \citenamefont
  {Girvin}(2003{\natexlab{a}})}]{Mitra:2003p1}%
  \BibitemOpen
  \bibfield  {author} {\bibinfo {author} {\bibfnamefont {A.}~\bibnamefont
  {Mitra}}\ and\ \bibinfo {author} {\bibfnamefont {S.~M.}\ \bibnamefont
  {Girvin}},\ }\bibfield  {title} {\bibinfo {title} {Electron/nuclear spin
  domain walls in quantum hall systems},\ }\href
  {https://doi.org/10.1103/PhysRevB.67.245311} {\bibfield  {journal} {\bibinfo
  {journal} {Phys. Rev. B}\ }\textbf {\bibinfo {volume} {67}},\ \bibinfo
  {pages} {245311} (\bibinfo {year} {2003}{\natexlab{a}})}\BibitemShut
  {NoStop}%
\bibitem [{\citenamefont {Abanin}\ \emph {et~al.}(2010)\citenamefont {Abanin},
  \citenamefont {Parameswaran}, \citenamefont {Kivelson},\ and\ \citenamefont
  {Sondhi}}]{NematicValleyPRB2010}%
  \BibitemOpen
  \bibfield  {author} {\bibinfo {author} {\bibfnamefont {D.~A.}\ \bibnamefont
  {Abanin}}, \bibinfo {author} {\bibfnamefont {S.~A.}\ \bibnamefont
  {Parameswaran}}, \bibinfo {author} {\bibfnamefont {S.~A.}\ \bibnamefont
  {Kivelson}},\ and\ \bibinfo {author} {\bibfnamefont {S.~L.}\ \bibnamefont
  {Sondhi}},\ }\bibfield  {title} {\bibinfo {title} {Nematic valley ordering in
  quantum hall systems},\ }\href {https://doi.org/10.1103/PhysRevB.82.035428}
  {\bibfield  {journal} {\bibinfo  {journal} {Phys. Rev. B}\ }\textbf {\bibinfo
  {volume} {82}},\ \bibinfo {pages} {035428} (\bibinfo {year}
  {2010})}\BibitemShut {NoStop}%
\bibitem [{\citenamefont {Kumar}\ \emph {et~al.}(2013)\citenamefont {Kumar},
  \citenamefont {Parameswaran},\ and\ \citenamefont {Sondhi}}]{Kumar}%
  \BibitemOpen
  \bibfield  {author} {\bibinfo {author} {\bibfnamefont {A.}~\bibnamefont
  {Kumar}}, \bibinfo {author} {\bibfnamefont {S.~A.}\ \bibnamefont
  {Parameswaran}},\ and\ \bibinfo {author} {\bibfnamefont {S.~L.}\ \bibnamefont
  {Sondhi}},\ }\bibfield  {title} {\bibinfo {title} {Microscopic theory of a
  quantum hall ising nematic: Domain walls and disorder},\ }\href
  {https://doi.org/10.1103/PhysRevB.88.045133} {\bibfield  {journal} {\bibinfo
  {journal} {Phys. Rev. B}\ }\textbf {\bibinfo {volume} {88}},\ \bibinfo
  {pages} {045133} (\bibinfo {year} {2013})}\BibitemShut {NoStop}%
\bibitem [{\citenamefont {Agarwal}\ \emph {et~al.}(2019)\citenamefont
  {Agarwal}, \citenamefont {Randeria}, \citenamefont {Yazdani}, \citenamefont
  {Sondhi},\ and\ \citenamefont {Parameswaran}}]{NematicDWFullTheory}%
  \BibitemOpen
  \bibfield  {author} {\bibinfo {author} {\bibfnamefont {K.}~\bibnamefont
  {Agarwal}}, \bibinfo {author} {\bibfnamefont {M.~T.}\ \bibnamefont
  {Randeria}}, \bibinfo {author} {\bibfnamefont {A.}~\bibnamefont {Yazdani}},
  \bibinfo {author} {\bibfnamefont {S.~L.}\ \bibnamefont {Sondhi}},\ and\
  \bibinfo {author} {\bibfnamefont {S.~A.}\ \bibnamefont {Parameswaran}},\
  }\bibfield  {title} {\bibinfo {title} {Topology- and symmetry-protected
  domain wall conduction in quantum hall nematics},\ }\href
  {https://doi.org/10.1103/PhysRevB.100.165103} {\bibfield  {journal} {\bibinfo
   {journal} {Phys. Rev. B}\ }\textbf {\bibinfo {volume} {100}},\ \bibinfo
  {pages} {165103} (\bibinfo {year} {2019})}\BibitemShut {NoStop}%
\bibitem [{\citenamefont {Danon}\ \emph {et~al.}(2019)\citenamefont {Danon},
  \citenamefont {Balram}, \citenamefont {S\'anchez},\ and\ \citenamefont
  {Rudner}}]{DanonIsingDW}%
  \BibitemOpen
  \bibfield  {author} {\bibinfo {author} {\bibfnamefont {J.}~\bibnamefont
  {Danon}}, \bibinfo {author} {\bibfnamefont {A.~C.}\ \bibnamefont {Balram}},
  \bibinfo {author} {\bibfnamefont {S.}~\bibnamefont {S\'anchez}},\ and\
  \bibinfo {author} {\bibfnamefont {M.~S.}\ \bibnamefont {Rudner}},\ }\bibfield
   {title} {\bibinfo {title} {Charge and spin textures of ising quantum hall
  ferromagnet domain walls},\ }\href
  {https://doi.org/10.1103/PhysRevB.100.235406} {\bibfield  {journal} {\bibinfo
   {journal} {Phys. Rev. B}\ }\textbf {\bibinfo {volume} {100}},\ \bibinfo
  {pages} {235406} (\bibinfo {year} {2019})}\BibitemShut {NoStop}%
\bibitem [{\citenamefont {Randeria}\ \emph {et~al.}(2019)\citenamefont
  {Randeria}, \citenamefont {Agarwal}, \citenamefont {Feldman}, \citenamefont
  {Ding}, \citenamefont {Ji}, \citenamefont {Cava}, \citenamefont {Sondhi},
  \citenamefont {Parameswaran},\ and\ \citenamefont {Yazdani}}]{NematicDWExpt}%
  \BibitemOpen
  \bibfield  {author} {\bibinfo {author} {\bibfnamefont {M.~T.}\ \bibnamefont
  {Randeria}}, \bibinfo {author} {\bibfnamefont {K.}~\bibnamefont {Agarwal}},
  \bibinfo {author} {\bibfnamefont {B.~E.}\ \bibnamefont {Feldman}}, \bibinfo
  {author} {\bibfnamefont {H.}~\bibnamefont {Ding}}, \bibinfo {author}
  {\bibfnamefont {H.}~\bibnamefont {Ji}}, \bibinfo {author} {\bibfnamefont
  {R.~J.}\ \bibnamefont {Cava}}, \bibinfo {author} {\bibfnamefont {S.~L.}\
  \bibnamefont {Sondhi}}, \bibinfo {author} {\bibfnamefont {S.~A.}\
  \bibnamefont {Parameswaran}},\ and\ \bibinfo {author} {\bibfnamefont
  {A.}~\bibnamefont {Yazdani}},\ }\bibfield  {title} {\bibinfo {title}
  {Interacting multi-channel topological boundary modes in a quantum hall
  valley system},\ }\href {https://doi.org/10.1038/s41586-019-0913-0}
  {\bibfield  {journal} {\bibinfo  {journal} {Nature}\ }\textbf {\bibinfo
  {volume} {566}},\ \bibinfo {pages} {363} (\bibinfo {year}
  {2019})}\BibitemShut {NoStop}%
\bibitem [{\citenamefont {Tarnopolsky}\ \emph {et~al.}(2019)\citenamefont
  {Tarnopolsky}, \citenamefont {Kruchkov},\ and\ \citenamefont
  {Vishwanath}}]{ChiralLimit}%
  \BibitemOpen
  \bibfield  {author} {\bibinfo {author} {\bibfnamefont {G.}~\bibnamefont
  {Tarnopolsky}}, \bibinfo {author} {\bibfnamefont {A.~J.}\ \bibnamefont
  {Kruchkov}},\ and\ \bibinfo {author} {\bibfnamefont {A.}~\bibnamefont
  {Vishwanath}},\ }\bibfield  {title} {\bibinfo {title} {Origin of magic angles
  in twisted bilayer graphene},\ }\href
  {https://doi.org/10.1103/PhysRevLett.122.106405} {\bibfield  {journal}
  {\bibinfo  {journal} {Phys. Rev. Lett.}\ }\textbf {\bibinfo {volume} {122}},\
  \bibinfo {pages} {106405} (\bibinfo {year} {2019})}\BibitemShut {NoStop}%
\bibitem [{\citenamefont {Khalaf}\ \emph
  {et~al.}(2020{\natexlab{a}})\citenamefont {Khalaf}, \citenamefont
  {Chatterjee}, \citenamefont {Bultinck}, \citenamefont {Zaletel},\ and\
  \citenamefont {Vishwanath}}]{khalaf2020charged}%
  \BibitemOpen
  \bibfield  {author} {\bibinfo {author} {\bibfnamefont {E.}~\bibnamefont
  {Khalaf}}, \bibinfo {author} {\bibfnamefont {S.}~\bibnamefont {Chatterjee}},
  \bibinfo {author} {\bibfnamefont {N.}~\bibnamefont {Bultinck}}, \bibinfo
  {author} {\bibfnamefont {M.~P.}\ \bibnamefont {Zaletel}},\ and\ \bibinfo
  {author} {\bibfnamefont {A.}~\bibnamefont {Vishwanath}},\ }\href@noop {}
  {\bibinfo {title} {Charged skyrmions and topological origin of
  superconductivity in magic angle graphene}} (\bibinfo {year}
  {2020}{\natexlab{a}}),\ \Eprint {https://arxiv.org/abs/2004.00638}
  {arXiv:2004.00638 [cond-mat.str-el]} \BibitemShut {NoStop}%
\bibitem [{\citenamefont {Khalaf}\ \emph
  {et~al.}(2020{\natexlab{b}})\citenamefont {Khalaf}, \citenamefont {Bultinck},
  \citenamefont {Vishwanath},\ and\ \citenamefont {Zaletel}}]{khalaf2020soft}%
  \BibitemOpen
  \bibfield  {author} {\bibinfo {author} {\bibfnamefont {E.}~\bibnamefont
  {Khalaf}}, \bibinfo {author} {\bibfnamefont {N.}~\bibnamefont {Bultinck}},
  \bibinfo {author} {\bibfnamefont {A.}~\bibnamefont {Vishwanath}},\ and\
  \bibinfo {author} {\bibfnamefont {M.~P.}\ \bibnamefont {Zaletel}},\
  }\href@noop {} {\bibinfo {title} {Soft modes in magic angle twisted bilayer
  graphene}} (\bibinfo {year} {2020}{\natexlab{b}}),\ \Eprint
  {https://arxiv.org/abs/2009.14827} {arXiv:2009.14827 [cond-mat.str-el]}
  \BibitemShut {NoStop}%
\bibitem [{\citenamefont {Mitra}\ and\ \citenamefont
  {Girvin}(2003{\natexlab{b}})}]{Mitra_Girvin}%
  \BibitemOpen
  \bibfield  {author} {\bibinfo {author} {\bibfnamefont {A.}~\bibnamefont
  {Mitra}}\ and\ \bibinfo {author} {\bibfnamefont {S.~M.}\ \bibnamefont
  {Girvin}},\ }\bibfield  {title} {\bibinfo {title} {Electron/nuclear spin
  domain walls in quantum hall systems},\ }\href
  {https://doi.org/10.1103/PhysRevB.67.245311} {\bibfield  {journal} {\bibinfo
  {journal} {Phys. Rev. B}\ }\textbf {\bibinfo {volume} {67}},\ \bibinfo
  {pages} {245311} (\bibinfo {year} {2003}{\natexlab{b}})}\BibitemShut
  {NoStop}%
\bibitem [{\citenamefont {Nam}\ and\ \citenamefont {Koshino}(2017)}]{nam2017}%
  \BibitemOpen
  \bibfield  {author} {\bibinfo {author} {\bibfnamefont {N.~N.~T.}\
  \bibnamefont {Nam}}\ and\ \bibinfo {author} {\bibfnamefont {M.}~\bibnamefont
  {Koshino}},\ }\bibfield  {title} {\bibinfo {title} {Lattice relaxation and
  energy band modulation in twisted bilayer graphene},\ }\href
  {https://doi.org/10.1103/PhysRevB.96.075311} {\bibfield  {journal} {\bibinfo
  {journal} {Phys. Rev. B}\ }\textbf {\bibinfo {volume} {96}},\ \bibinfo
  {pages} {075311} (\bibinfo {year} {2017})}\BibitemShut {NoStop}%
\bibitem [{\citenamefont {Carr}\ \emph {et~al.}(2019)\citenamefont {Carr},
  \citenamefont {Fang}, \citenamefont {Zhu},\ and\ \citenamefont
  {Kaxiras}}]{carr2019}%
  \BibitemOpen
  \bibfield  {author} {\bibinfo {author} {\bibfnamefont {S.}~\bibnamefont
  {Carr}}, \bibinfo {author} {\bibfnamefont {S.}~\bibnamefont {Fang}}, \bibinfo
  {author} {\bibfnamefont {Z.}~\bibnamefont {Zhu}},\ and\ \bibinfo {author}
  {\bibfnamefont {E.}~\bibnamefont {Kaxiras}},\ }\bibfield  {title} {\bibinfo
  {title} {Exact continuum model for low-energy electronic states of twisted
  bilayer graphene},\ }\href {https://doi.org/10.1103/PhysRevResearch.1.013001}
  {\bibfield  {journal} {\bibinfo  {journal} {Phys. Rev. Research}\ }\textbf
  {\bibinfo {volume} {1}},\ \bibinfo {pages} {013001} (\bibinfo {year}
  {2019})}\BibitemShut {NoStop}%
\bibitem [{\citenamefont {Qi}(2011)}]{Qi}%
  \BibitemOpen
  \bibfield  {author} {\bibinfo {author} {\bibfnamefont {X.-L.}\ \bibnamefont
  {Qi}},\ }\bibfield  {title} {\bibinfo {title} {Generic wave-function
  description of fractional quantum anomalous hall states and fractional
  topological insulators},\ }\href
  {https://doi.org/10.1103/PhysRevLett.107.126803} {\bibfield  {journal}
  {\bibinfo  {journal} {Phys. Rev. Lett.}\ }\textbf {\bibinfo {volume} {107}},\
  \bibinfo {pages} {126803} (\bibinfo {year} {2011})}\BibitemShut {NoStop}%
\bibitem [{\citenamefont {Scaffidi}\ and\ \citenamefont
  {M\"oller}(2012)}]{Scaffidi}%
  \BibitemOpen
  \bibfield  {author} {\bibinfo {author} {\bibfnamefont {T.}~\bibnamefont
  {Scaffidi}}\ and\ \bibinfo {author} {\bibfnamefont {G.}~\bibnamefont
  {M\"oller}},\ }\bibfield  {title} {\bibinfo {title} {Adiabatic continuation
  of fractional chern insulators to fractional quantum hall states},\ }\href
  {https://doi.org/10.1103/PhysRevLett.109.246805} {\bibfield  {journal}
  {\bibinfo  {journal} {Phys. Rev. Lett.}\ }\textbf {\bibinfo {volume} {109}},\
  \bibinfo {pages} {246805} (\bibinfo {year} {2012})}\BibitemShut {NoStop}%
\bibitem [{\citenamefont {Wu}\ \emph {et~al.}(2012)\citenamefont {Wu},
  \citenamefont {Regnault},\ and\ \citenamefont {Bernevig}}]{Regnault}%
  \BibitemOpen
  \bibfield  {author} {\bibinfo {author} {\bibfnamefont {Y.-L.}\ \bibnamefont
  {Wu}}, \bibinfo {author} {\bibfnamefont {N.}~\bibnamefont {Regnault}},\ and\
  \bibinfo {author} {\bibfnamefont {B.~A.}\ \bibnamefont {Bernevig}},\
  }\bibfield  {title} {\bibinfo {title} {Gauge-fixed wannier wave functions for
  fractional topological insulators},\ }\href
  {https://doi.org/10.1103/PhysRevB.86.085129} {\bibfield  {journal} {\bibinfo
  {journal} {Phys. Rev. B}\ }\textbf {\bibinfo {volume} {86}},\ \bibinfo
  {pages} {085129} (\bibinfo {year} {2012})}\BibitemShut {NoStop}%
\bibitem [{\citenamefont {Barkeshli}\ and\ \citenamefont {Qi}(2012)}]{QiPRX}%
  \BibitemOpen
  \bibfield  {author} {\bibinfo {author} {\bibfnamefont {M.}~\bibnamefont
  {Barkeshli}}\ and\ \bibinfo {author} {\bibfnamefont {X.-L.}\ \bibnamefont
  {Qi}},\ }\bibfield  {title} {\bibinfo {title} {Topological nematic states and
  non-abelian lattice dislocations},\ }\href
  {https://doi.org/10.1103/PhysRevX.2.031013} {\bibfield  {journal} {\bibinfo
  {journal} {Phys. Rev. X}\ }\textbf {\bibinfo {volume} {2}},\ \bibinfo {pages}
  {031013} (\bibinfo {year} {2012})}\BibitemShut {NoStop}%
\bibitem [{\citenamefont {Liu}\ \emph {et~al.}(2019{\natexlab{b}})\citenamefont
  {Liu}, \citenamefont {Liu},\ and\ \citenamefont {Dai}}]{Liu}%
  \BibitemOpen
  \bibfield  {author} {\bibinfo {author} {\bibfnamefont {J.}~\bibnamefont
  {Liu}}, \bibinfo {author} {\bibfnamefont {J.}~\bibnamefont {Liu}},\ and\
  \bibinfo {author} {\bibfnamefont {X.}~\bibnamefont {Dai}},\ }\bibfield
  {title} {\bibinfo {title} {Pseudo landau level representation of twisted
  bilayer graphene: Band topology and implications on the correlated insulating
  phase},\ }\href {https://doi.org/10.1103/PhysRevB.99.155415} {\bibfield
  {journal} {\bibinfo  {journal} {Phys. Rev. B}\ }\textbf {\bibinfo {volume}
  {99}},\ \bibinfo {pages} {155415} (\bibinfo {year}
  {2019}{\natexlab{b}})}\BibitemShut {NoStop}%
\bibitem [{\citenamefont {Li}\ \emph {et~al.}(2014)\citenamefont {Li},
  \citenamefont {Zhang}, \citenamefont {Niu},\ and\ \citenamefont
  {MacDonald}}]{Li2014}%
  \BibitemOpen
  \bibfield  {author} {\bibinfo {author} {\bibfnamefont {X.}~\bibnamefont
  {Li}}, \bibinfo {author} {\bibfnamefont {F.}~\bibnamefont {Zhang}}, \bibinfo
  {author} {\bibfnamefont {Q.}~\bibnamefont {Niu}},\ and\ \bibinfo {author}
  {\bibfnamefont {A.~H.}\ \bibnamefont {MacDonald}},\ }\bibfield  {title}
  {\bibinfo {title} {Spontaneous layer-pseudospin domain walls in bilayer
  graphene},\ }\href {https://doi.org/10.1103/PhysRevLett.113.116803}
  {\bibfield  {journal} {\bibinfo  {journal} {Phys. Rev. Lett.}\ }\textbf
  {\bibinfo {volume} {113}},\ \bibinfo {pages} {116803} (\bibinfo {year}
  {2014})}\BibitemShut {NoStop}%
\bibitem [{\citenamefont {He}\ \emph {et~al.}(2020)\citenamefont {He},
  \citenamefont {Goldhaber-Gordon},\ and\ \citenamefont
  {Law}}]{He_MagReversal}%
  \BibitemOpen
  \bibfield  {author} {\bibinfo {author} {\bibfnamefont {W.-Y.}\ \bibnamefont
  {He}}, \bibinfo {author} {\bibfnamefont {D.}~\bibnamefont
  {Goldhaber-Gordon}},\ and\ \bibinfo {author} {\bibfnamefont {K.~T.}\
  \bibnamefont {Law}},\ }\bibfield  {title} {\bibinfo {title} {Giant orbital
  magnetoelectric effect and current-induced magnetization switching in twisted
  bilayer graphene},\ }\href {https://doi.org/10.1038/s41467-020-15473-9}
  {\bibfield  {journal} {\bibinfo  {journal} {Nature Communications}\ }\textbf
  {\bibinfo {volume} {11}},\ \bibinfo {pages} {1650} (\bibinfo {year}
  {2020})}\BibitemShut {NoStop}%
\bibitem [{\citenamefont {Huang}\ \emph {et~al.}(2020)\citenamefont {Huang},
  \citenamefont {Wei},\ and\ \citenamefont {MacDoanld}}]{huang2020current}%
  \BibitemOpen
  \bibfield  {author} {\bibinfo {author} {\bibfnamefont {C.}~\bibnamefont
  {Huang}}, \bibinfo {author} {\bibfnamefont {N.}~\bibnamefont {Wei}},\ and\
  \bibinfo {author} {\bibfnamefont {A.}~\bibnamefont {MacDoanld}},\ }\href@noop
  {} {\bibinfo {title} {Current driven magnetization reversal in orbital chern
  insulators}} (\bibinfo {year} {2020}),\ \Eprint
  {https://arxiv.org/abs/2007.05990} {arXiv:2007.05990 [cond-mat.mes-hall]}
  \BibitemShut {NoStop}%
\bibitem [{\citenamefont {Chalker}\ and\ \citenamefont
  {Coddington}(1988)}]{Chalker_1988}%
  \BibitemOpen
  \bibfield  {author} {\bibinfo {author} {\bibfnamefont {J.~T.}\ \bibnamefont
  {Chalker}}\ and\ \bibinfo {author} {\bibfnamefont {P.~D.}\ \bibnamefont
  {Coddington}},\ }\bibfield  {title} {\bibinfo {title} {Percolation, quantum
  tunnelling and the integer hall effect},\ }\href
  {https://doi.org/10.1088/0022-3719/21/14/008} {\bibfield  {journal} {\bibinfo
   {journal} {Journal of Physics C: Solid State Physics}\ }\textbf {\bibinfo
  {volume} {21}},\ \bibinfo {pages} {2665} (\bibinfo {year}
  {1988})}\BibitemShut {NoStop}%
\bibitem [{\citenamefont {Hejazi}\ \emph {et~al.}(2020)\citenamefont {Hejazi},
  \citenamefont {Chen},\ and\ \citenamefont {Balents}}]{hejazi2020hybrid}%
  \BibitemOpen
  \bibfield  {author} {\bibinfo {author} {\bibfnamefont {K.}~\bibnamefont
  {Hejazi}}, \bibinfo {author} {\bibfnamefont {X.}~\bibnamefont {Chen}},\ and\
  \bibinfo {author} {\bibfnamefont {L.}~\bibnamefont {Balents}},\ }\href@noop
  {} {\bibinfo {title} {Hybrid wannier chern bands in magic angle twisted
  bilayer graphene and the quantized anomalous hall effect}} (\bibinfo {year}
  {2020}),\ \Eprint {https://arxiv.org/abs/2007.00134} {arXiv:2007.00134
  [cond-mat.mes-hall]} \BibitemShut {NoStop}%
\bibitem [{\citenamefont {Kang}\ and\ \citenamefont {Vafek}(2020)}]{kang2020}%
  \BibitemOpen
  \bibfield  {author} {\bibinfo {author} {\bibfnamefont {J.}~\bibnamefont
  {Kang}}\ and\ \bibinfo {author} {\bibfnamefont {O.}~\bibnamefont {Vafek}},\
  }\bibfield  {title} {\bibinfo {title} {Non-abelian dirac node braiding and
  near-degeneracy of correlated phases at odd integer filling in magic-angle
  twisted bilayer graphene},\ }\href
  {https://doi.org/10.1103/PhysRevB.102.035161} {\bibfield  {journal} {\bibinfo
   {journal} {Phys. Rev. B}\ }\textbf {\bibinfo {volume} {102}},\ \bibinfo
  {pages} {035161} (\bibinfo {year} {2020})}\BibitemShut {NoStop}%
\bibitem [{\citenamefont {Soejima}\ \emph {et~al.}(2020)\citenamefont
  {Soejima}, \citenamefont {Parker}, \citenamefont {Bultinck}, \citenamefont
  {Hauschild},\ and\ \citenamefont {Zaletel}}]{soejima2020efficient}%
  \BibitemOpen
  \bibfield  {author} {\bibinfo {author} {\bibfnamefont {T.}~\bibnamefont
  {Soejima}}, \bibinfo {author} {\bibfnamefont {D.~E.}\ \bibnamefont {Parker}},
  \bibinfo {author} {\bibfnamefont {N.}~\bibnamefont {Bultinck}}, \bibinfo
  {author} {\bibfnamefont {J.}~\bibnamefont {Hauschild}},\ and\ \bibinfo
  {author} {\bibfnamefont {M.~P.}\ \bibnamefont {Zaletel}},\ }\href@noop {}
  {\bibinfo {title} {Efficient simulation of moire materials using the density
  matrix renormalization group}} (\bibinfo {year} {2020}),\ \Eprint
  {https://arxiv.org/abs/2009.02354} {arXiv:2009.02354 [cond-mat.str-el]}
  \BibitemShut {NoStop}%
\bibitem [{Note1()}]{Note1}%
  \BibitemOpen
  \bibinfo {note} {We thank Allan MacDonald for suggesting this choice of
  reference projector.}\BibitemShut {Stop}%
\end{thebibliography}
\end{document}